\documentclass[12pt]{article}

\ifx\pdfoutput\undefined
\usepackage[dvips,bookmarks]{hyperref}
\else
\usepackage{hyperref}
\fi
\hypersetup{colorlinks=false,bookmarksopen,bookmarksnumbered,citecolor=blue,
   pdfstartview=FitH}

\usepackage{latexsym}
\usepackage{amssymb,amsfonts,amsmath}
\usepackage{graphicx} 
\usepackage{indentfirst}
\usepackage{bbm}

\parskip=\medskipamount

\newcommand {\cD}{{\cal D}}
\newcommand {\cE}{{\cal E}}
\newcommand {\cF}{{\cal F}}

\newcommand {\cK}{{\cal K}}
\newcommand {\cL}{{\cal L}}

\newcommand {\cN}{{\cal N}}

\newcommand {\cS}{{\cal S}}

\newcommand {\cW}{{\cal W}}

\def\a{\alpha}
\def\b{\beta}

\def\d{\delta}

\def\f{\phi}
\def\g{\gamma}
\def\G{\Gamma}

\def\j{\psi}

\def\l{\lambda}
\def\m{\mu}

\def\s{\sigma}

\def\D{\Delta}

\def\S{\Sigma}

\def\ri{{\rm i}}
\def\re{{\rm e}}

\newcommand{\ad}{{\dot{\alpha}}}                           
\newcommand{\bd}{{\dot{\beta}}}                            
\newcommand{\ve}{\varepsilon}                            
\newcommand{\cDB}{{\bar\cD}}                            

\newcommand{\hf}{\frac12}

\newcommand{\vf}{\varphi}

\newcommand{\be}{\begin{equation}}
\newcommand{\ee}{\end{equation}}
\newcommand{\bea}{\begin{eqnarray}}
\newcommand{\eea}{\end{eqnarray}}
\newcommand{\non}{\nonumber}
\newcommand{\ba}{\begin{array}}
\newcommand{\ea}{\end{array}}

\newcommand{\bm}[1]{\mbox{\boldmath$#1$}}

\def\double #1{#1{\hbox{\kern-2pt $#1$}}}

\newcommand{\gd}{{\dot\g}}

\newcommand{\bsubeq}{\begin{subequations}}
\newcommand{\esubeq}{\end{subequations}}

\newcommand{\eps}{\ve}

\newcommand{\dalpha}{{\dot{\alpha}}}
\newcommand{\dbeta}{{\dot{\beta}}}
\newcommand{\dgamma}{{\dot{\gamma}}}

\newcommand{\N}{{\mathcal N}}
\newcommand{\eol}{\notag \\}
\newcommand{\rd}{\mathrm d}
\newcommand{\CD}{\cD}
\newcommand{\BCD}{{\bar\cD}}

\begin{document}

\begin{titlepage}
\begin{flushright}
June 2011\\
\end{flushright}
\vspace{5mm}

\begin{center}
{\Large \bf A dual formulation of supergravity-matter theories}
\\ 
\end{center}

\begin{center}

{\bf Daniel Butter and  Sergei M. Kuzenko }

\footnotesize{
{\it School of Physics M013, The University of Western Australia\\
35 Stirling Highway, Crawley W.A. 6009, Australia}}  ~\\
\texttt{dbutter,\,kuzenko@cyllene.uwa.edu.au}\\
\vspace{2mm}

\end{center}
\vspace{5mm}

\begin{abstract}
\baselineskip=14pt

Generating supersymmetric AdS solutions in non-minimal supergravity in four
dimensions is notoriously difficult. Indeed, it is a longstanding lore that such
solutions exist \emph{only} for old minimal supergravity. In this paper,
we construct a dual formulation for general $\N=1$ supergravity-matter
systems that avoids the problem.
In the case of pure supergravity without a cosmological constant,
it coincides with the usual non-minimal ($n=-1$) supergravity, but in
the presence of matter (or a cosmological constant) our formulation
differs considerably. We also elaborate upon the framework of
conformal superspace and the compensator method as applied to our theory. In
particular, we show that one can encode the details of the K\"ahler potential 
and superpotential entirely within the geometry of superspace so that the
general sigma-model action is encoded in a single compact term: the supervolume.
Finally, we discuss the issue of supercurrents and propose a general form
for the supercurrent in AdS.

\end{abstract}
\vspace{0.5cm}


\vfill
\end{titlepage}

\newpage
\renewcommand{\thefootnote}{\arabic{footnote}}
\setcounter{footnote}{0}

\tableofcontents


\numberwithin{equation}{section}



\newcommand{\fint}{{\int \rd^4x\, \rd^4\theta\, E\, }}
\newcommand{\cint}{{\int \rd^4x\, \rd^2\theta\, \cE\, }}
\newcommand{\acint}{{\int \rd^4x\, \rd^2\bar\theta\, \bar \cE\, }}
\newcommand{\compint}{{\int \rd^4x\, e\, }}

\newcommand{\hfint}{{\int \rd^4x\, \rd^4\theta\, \hat E\, }}
\newcommand{\hcint}{{\int \rd^4x\, \rd^2\theta\, \hat\cE\, }}
\newcommand{\hacint}{{\int \rd^4x\, \rd^2\bar\theta\, \bar {\hat\cE}\, }}

\renewcommand{\D}{\mathbf D}
\newcommand{\A}{\mathbf A}
\newcommand{\K}{\mathbf K}

\newcommand{\cUCD}{{\hat\cD}}
\newcommand{\cUBCD}{\bar{\hat\cD}}
\newcommand{\cUR}{{\hat R}}
\newcommand{\cUG}{{\hat G}}
\newcommand{\cUX}{{\hat X}}

\newcommand{\UCD}{{\bm \cD}}
\newcommand{\UBCD}{{\bar \UCD}}
\newcommand{\UR}{{\mathbf R}}
\newcommand{\UG}{{\mathbf G}}
\newcommand{\UX}{{\mathbf X}}
\newcommand{\UW}{{\mathbf W}}
\newcommand{\UF}{{\bm \cF}}
\newcommand{\UL}{{\bm \cL}}
\newcommand{\UE}{{\mathbf E}}
\newcommand{\UPsi}{{\bm \Psi}}

\newcommand{\FCD}{{\mathbb D}}
\newcommand{\FBCD}{{\bar \FCD}}
\newcommand{\FR}{{\mathbb R}}
\newcommand{\FG}{{\mathbb G}}
\newcommand{\FW}{{\mathbb W}}
\newcommand{\FT}{{\mathbb T}}
\newcommand{\FF}{{\mathbb F}}
\newcommand{\FK}{{\mathbb K}}

\newcommand{\cFCD}{\hat \FCD}
\newcommand{\cFBCD}{\hat \FBCD}
\newcommand{\cFR}{\hat \FR}
\newcommand{\cFG}{\hat \FG}
\newcommand{\cFT}{\hat \FT}

\section{Introduction}

In a recent paper \cite{BK2011} we have shown that the structure of consistent 
supercurrents in four-dimensional $\cN=1$ anti-de Sitter (AdS) supersymmetry 
considerably differs from that in the Minkowski case. There are only two
irreducible AdS supercurrents, with $12+12$ and $20+20$ degrees of freedom.
The former is naturally associated with the so-called longitudinal action
$S^{||}_{(3/2)}$ for a massless superspin-3/2 multiplet, which involves a real superfield
$H_{\alpha \dalpha}$ and a chiral superfield $\sigma$. The latter is associated
with a unique dual formulation $S^{\perp}_{(3/2)}$ where the chiral superfield
is replaced by a complex linear superfield $\Gamma$. Both these actions
represent the lowest superspin limits of two infinite series of dual models for off-shell massless
higher spin supermultiplets in AdS
 \cite{KS94}. (The actions $S^{||}_{(3/2)}$ and $S^{\perp}_{(3/2)}$
coincide, respectively, with the actions $S_{\rm old}$ and $S_{n=-1}$ referred
to in the introduction of \cite{BK2011}. Explicit expressions for each are
given in Appendix A.)

The longitudinal formulation $S^{||}_{3/2}$ has a natural interpretation in
supergravity. It results from the linearization around the AdS background
of old minimal $(n=-1/3)$ supergravity with a cosmological term.
One can show that in the Minkowski limit, $S^{\perp}_{(3/2)}$ arises as a
linearization of non-minimal $(n=-1)$ supergravity.\footnote{See \cite{GGRS,BK}
for reviews of different off-shell versions of $\cN=1$ supergravity}
However, there exists a general belief that $S^{\perp}_{(3/2)}$ cannot be obtained 
by linearizing a supergravity action around a supersymmetric AdS solution. This belief can be traced back  to 1983
when it was pointed out in \cite{GGRS}  that only old minimal supergravity admits a well-defined AdS
solution.\footnote{Specifically, on page 336 of {\it Superspace} \cite{GGRS} Grisaru {\it et al.}
stated the following. Except for the old minimal  formulation of $\cN=1$ supergravity, ``we find the
strange pathologies: de Sitter space cannot be described for $n\neq -1/3$ in a globally  (de Sitter)
supersymmetric way.''} This point was further elaborated upon in \cite{DG}.

In the present paper, we propose a novel formulation for $n=-1$ non-minimal supergravity 
and its matter couplings which resolves this puzzle. This formulation provides a simple description
of supergravity with a cosmological term such that (i) $\cN=1$ AdS superspace is its 
maximally symmetric solution;  and (ii)  $S^{\perp}_{(3/2)}$ is the linearized action around
the AdS background. We also explain how we get around the analysis
given in \cite{GGRS}.

This paper is organized as follows. In section 2, we describe a globally
supersymmetric duality between a chiral multiplet and a complex linear multiplet
with a modified constraint. We utilize this simple model in section 3 to
describe a new set of matter couplings in non-minimal supergravity. In section 4,
we specialize to the case of pure supergravity with a cosmological constant,
demonstrate that AdS is indeed a solution, and describe the way around the ``no-go''
analysis of \cite{GGRS}. In the remainder of the paper we elaborate upon
this construction: in section 5, we discuss in some generality the method of compensators
in conformal supergravity and in section 6, we describe a method to ``geometrize''
the new matter couplings in the spirit of K\"ahler
superspace \cite{Binetruy:1987qw, Binetruy:2000zx}. We conclude by briefly
discussing supercurrents and consider the case of a non-linear sigma model
in AdS. A brief appendix is included discussing the linearized supergravity
actions in AdS. Throughout this paper, we shall use the superspace conventions
of \cite{BK}.\footnote{These conventions are nearly identical to those
of Wess and Bagger \cite{WB}. To convert our notation to theirs, one
replaces $R \rightarrow 2 R$, $G_{\alpha \dalpha} \rightarrow 2 G_{\alpha \dalpha}$, and
$W_{\alpha \beta \gamma} \rightarrow 2 W_{\alpha \beta \gamma}$.
In addition, the vector derivative is defined by $\{\CD_\alpha, \BCD_\dalpha\} = -2\ri \CD_{\alpha \dalpha}$.}

\section{The improved complex linear multiplet}
Let us begin with a simple superconformal model in global supersymmetry
involving a chiral superfield $\phi$,
\begin{align}\label{2.1}
S = \int \rd^4x\, \rd^4\theta\, \phi \bar\phi~, \qquad {\bar D}_\ad \f = 0~.
\end{align}
The action is superconformal provided $\phi$ transforms with unit dilatation
weight and $U(1)_R$ weight $2/3$.\footnote{Our convention for the $U(1)_R$
weight is such that the derivative $D_\alpha$ has weight $-1$.} There exists a
well understood duality \cite{Zumino1980,GS81}
between this model and that involving a complex linear superfield $\S$
obeying the constraint 
\begin{align}\label{eq_CLconstraint}
\bar D^2 \S = 0 ~,\qquad D^2 \bar \S = 0~.
\end{align}
This constraint is conformally invariant provided $\S$ has dilatation
weight $\Delta$ and $U(1)_R$ weight $w$ related by
\begin{align}
2 \Delta - 3 w = 4~.
\end{align}
This condition is solved for a one-parameter family,\footnote{The parameter $n$ was first introduced
in the context of non-minimal supergravity \cite{SG}.}  
labelled by $n$,
with
\begin{align}\label{eq_CLweights}
\Delta = \frac{2}{3n + 1}~, \qquad w = -\frac{4n}{3n + 1}~.
\end{align}
For $n\neq 0, -1/3$, a duality between $\phi$ and $\S$ may be carried out.
Our concern here is the case $n=-1$. For that situation, we may introduce the
first-order action
\begin{align}\label{2.5}
S = \int \rd^4x\, \rd^4\theta\, \Big(\phi \bar\phi
	- \frac{1}{3} \S \phi^3 - \frac{1}{3} \bar\S \bar\phi^3\Big)
\end{align}
where $\phi$ is now an unconstrained complex superfield. The
constraint \eqref{eq_CLconstraint} is solved by
$\Sigma = \bar D_\dalpha \bar\psi^\dalpha$ for an unconstrained
superfield $\bar\psi^\dalpha$; therefore, the equation of motion for
$\S$ enforces the chirality of $\phi$ and then this action reduces to
(\ref{2.1}). Instead, we may solve the equation of motion for $\f$ to obtain
\begin{align}
\phi^3 = \S^{-2} \bar\S^{-1}~.
\end{align}
Plugging this in the action yields the dual formulation
\begin{align}
S_{\rm dual} = \frac{1}{3} \int \rd^4x\, \rd^4\theta\, (\S \bar\S)^{-1}~.
\end{align}

Now let us consider an extension of the action (\ref{2.1}) by introducing the
only possible superconformal interaction term for $\phi$:
\begin{align} \label{2.8}
S = \int \rd^4x\, \rd^4\theta\, \phi \bar\phi
	-\frac{\mu}{3} \int \rd^4x\, \rd^2\theta\, \phi^3
	-\frac{\bar\mu}{3} \int \rd^4x\, \rd^2\bar\theta\, \bar\phi^3
\end{align}
where $\mu$ is a complex parameter. It is easy to see that this arises
from the same first order action
\begin{align}\label{2.9}
S = \int \rd^4x\, \rd^4\theta\, \Big(\phi \bar\phi
	- \frac{1}{3} \Gamma \phi^3 - \frac{1}{3} \bar\Gamma \bar\phi^3\Big)
\end{align}
if we require $\Gamma$ to obey the \emph{improved} complex linear
constraint\footnote{It is of interest to point out that eq.  (\ref{2.10})
also emerges  as one of the constraints obeyed by the complex linear
Goldstino superfield  introduced in \cite{KT}.}
\begin{align} \label{2.10}
\bar D^2 \Gamma = -4 \mu~, \qquad D^2 \bar \Gamma = -4\bar\mu~.
\end{align}
This constraint is indeed superconformally invariant  since $\bar D^2 \Gamma$
possesses vanishing dilatation and $U(1)_R$ weights. The dual action has then exactly
the  same form as above, that is
\begin{align}
S_{\rm dual} = \frac{1}{3} \int \rd^4x\, \rd^4\theta\, (\Gamma \bar\Gamma)^{-1}~.
\end{align}
The constraint (\ref{2.10}) is responsible for generating the nontrivial interaction
terms.

\section{Matter couplings in non-minimal supergravity revisited}
We now consider a locally supersymmetric generalization of the action (\ref{2.8}) of the form
\begin{align}
S = -3 \fint \phi \bar\phi \,{\rm e}^{-K/3}
	+ \cint \phi^3 W
	+ \acint \bar\phi^3 \bar W~.
	\label{3.1}
\end{align}
This is the superconformal version (in the spirit of, e.g., Kugo and
Uehara \cite{Kugo}) of the most general non-linear sigma model action coupled
to supergravity (see, e.g. \cite{WB} for a review).
The K\"ahler potential,  $K= K(\vf^i, \bar \vf^{\bar j})$, is a real function 
of the covariantly chiral superfields $\vf^i$ and their conjugates
$\bar \vf^{\bar j}$, obeying $\bar \cD_\ad \vf^i =0$.
The superpotential,  $W = W(\vf^i)$,  is a holomorphic function of $\vf^i$ alone.
The conformal compensator $\phi$ is also a covariantly chiral scalar
superfield, $\bar \cD_\ad \f=0$, and it must be nowhere vanishing, $\f \neq 0$.
To describe off-shell supergravity, in this section we use the Wess-Zumino
superspace formulation \cite{WZ} (see \cite{BK,WB} for reviews), which at the
component level generates the old minimal supergravity multiplet \cite{old}.

The action (\ref{3.1}) is invariant under K\"ahler transformations,
\begin{align}
K   \rightarrow K + F + \bar F, \qquad W \rightarrow {\rm e}^{-F} \,W, \qquad
\phi \rightarrow {\rm e}^{F/3}\,\phi~,
\end{align}
with  $F(\vf^i)$ an arbitrary holomorphic function.
Moreover, the action is invariant under  super-Weyl transformations \cite{HT}
\begin{subequations} \label{superweyl}
\bea
\cD_\a &\to& {\rm e}^{ \s/2 - {\bar \s} } \Big(
\cD_\a - (\cD^\b \s) \, M_{\a \b} \Big) ~, \qquad {\bar \cD}_\ad \s =0 \\
\cDB_\ad & \to & {\rm e}^{ {\bar \s}/2 - \s } \Big(
\cDB_\ad -  (\cDB^\bd {\bar \s}) {\bar M}_{\bd\ad} \Big)~,
\eea
\end{subequations}
which act on the conformal compensator $\f$ and the matter fields $\vf^i$ as follows:
\bea
\f &\to& {\rm e}^{-\s} \f~,\label{3.4} \\
\vf^i &\to& \vf^i~.
\eea
The standard formulation of this model described, e.g., in \cite{WB}
may be derived from (\ref{3.1}) by enforcing the super-Weyl gauge choice $\phi = 1$. Then
every K\"ahler transformation must be accompanied by a compensating
super-Weyl transformation to maintain the gauge choice $\phi=1$.

We will now perform the very same type of duality transformation as in the
previous section. We introduce the first-order action
\begin{align}
S = \fint \Big(-3 \phi \bar\phi \,{\rm e}^{-K/3}
	+ \Gamma \phi^3 + \bar\Gamma \bar\phi^3\Big)
	\label{3.6}
\end{align}
in which $\f$ is an unconstrained scalar, while the complex scalar 
$\G$ obeys  the {\it improved} linear constraint
\begin{align}
-\frac{1}{4} (\bar \cD^2 - 4 R) \Gamma = W(\vf)~.
\label{3.7}
\end{align}
This constraint is model-dependent.\footnote{In global supersymmetry, constraints  
of the form (\ref{3.7}) were introduced for the first time by Deo and Gates \cite{DG85}.
In the context of supergravity, such constraints have recently been used in \cite{KT} to generate
couplings of the Goldstino superfield to chiral  matter.}
For $W =0$ it defines the standard complex 
linear multiplet. 

To maintain K\"ahler invariance of the action (\ref{3.6}), $\Gamma$ must transform as
\begin{align}
\Gamma \rightarrow {\rm e}^{-F} \Gamma~.
\end{align}
Since eq. (\ref{superweyl}) implies
\be
(\bar \cD^2 - 4 { R})  \to {\rm e}^{-2  \s} \,
(\bar \cD^2 - 4 { R})\,{\rm e}^{\bar  \s}
\label{quadr}
\ee
when acting on a scalar superfield, 
and since $W(\vf)$ is super-Weyl invariant, 
the super-Weyl transformation of $\G$ must be
\bea
\G &\to& {\rm e}^{2\s -\bar \s } \G~.
\label{3.10}
\eea

The model introduced above is equivalent to (\ref{3.1}). Indeed, let us first vary  (\ref{3.6})
with respect to $\G$ using 
\bea
\d \G = {\bar \cD}_\ad  \d \bar \j^\ad~, 
\eea
with $\d \bar \j^\ad$ an arbitrary spinor superfield. The equation of motion for $\G$ is equivalent to 
the chirality of  $ \f$, and then 
the first-order action (\ref{3.6}) turns into (\ref{3.1}), as a consequence of the standard identity 
\bea\label{eq_DtoFterm}
 \fint U=  -\frac{1}{4}\cint (\bar \cD^2 - 4 { R}) U
\eea
and the constraint (\ref{3.7}).
On the other hand, integrating out $\phi$ yields
\begin{align}
\phi^3 = {\rm e}^{-K} \Gamma^{-2} \bar\Gamma^{-1}~,
\end{align}
which upon insertion into (\ref{3.6}) gives the dual action
\begin{align}
S_{\rm dual} = -\int \rd^4x\, \rd^4\theta\, E\,{\rm e}^{-K} (\Gamma \bar\Gamma)^{-1}~,
\label{3.14}
\end{align}
which is K\"ahler invariant.
This action provides a novel formulation for general matter couplings
in $n=-1$ non-minimal supergravity. The superpotential is present in $S_{\rm dual} $ via the constraint
on $\Gamma$, eq. (\ref{3.7}).

At this point, it is natural to take a pause and recall
the traditional description of matter couplings in the non-minimal supergravity 
(see textbooks \cite{GGRS,BK} for more detailed presentations).
It is well known that  any $\cN=1$ supergravity-matter system, including the new minimal \cite{new}
 ($n=0$)
 and non-minimal \cite{non-min,SG} 
 $(n\neq -1/3, 0)$ supergravity theories, can be realized as a super-Weyl invariant coupling of 
 the old minimal supergravity to matter \cite{FGKV,BK}. 
 To describe the non-minimal supergravity in such a setting, we follow\footnote{A similar approach was
 developed by Kugo and Uehara \cite{Kugo2}.} 
 \cite{BKS,BK} 
and  couple the old minimal supergravity to a complex linear superfield $\S$ and its conjugate $\bar \S$, 
 \bea
-\frac{1}{4} (\bar \cD^2 - 4 R) \S=0 ~,
\label{3.15}
\eea
with the super-Weyl transformation law\footnote{Within the Wess-Zumino superspace
formulation, this transformation rule is equivalent to the dilatation and
$U(1)_R$ weights given in \eqref{eq_CLweights}.}
\bea
\S \to \exp \Big[ \frac{3n-1}{3n+1}\s - \bar \s\Big] \S~.
\label{3.16}
\eea
The  super-Weyl invariant action for pure non-minimal supergravity is
\bea
S_{\text{non-minimal}} = \frac{1}{n} \int \rd^4x\, \rd^4\theta\, E\, ( \S \, \bar\S)^{ (3n+1) /2 }~.
\eea
There is a well-elaborated scheme to generalize this action to incorporate matter couplings. 
To include a superpotential, for instance, we have to deform this action by adding appropriate terms, 
while keeping the constraint (\ref{3.15}) fixed. The latter feature is universal. 
${}$For general supergravity-matter couplings, the constraint (\ref{3.15}) remains the same.

In our formulation, the superpotential does not appear explicitly 
in the supergravity matter action (\ref{3.14}). 
The superpotential is instead incorporated by appropriately deforming 
the constraint on $\G$, eq. (\ref{3.7}).
By comparing the super-Weyl transformation laws (\ref{3.10}) and (\ref{3.16}), 
it is seen that our formulation corresponds to $n=-1$.

As shown in \cite{BK,BKS}, the geometric properties of non-minimal supergravity can 
be described in terms of super-Weyl-inert covariant derivatives, 
${\mathbb D}_A =( {\mathbb D}_a, {\mathbb D}_\a , {\bar{\mathbb D}}^\ad)$, 
constructed from the minimal ones, $\cD_A=(\cD_a , \cD_\a, \bar \cD^\ad)$, 
and a complex conformal compensator $\mathbb F$, which transforms
under the super-Weyl transformations as
\begin{align}
\mathbb F \rightarrow \mathbb F \,\re^{-\sigma / 2 + \bar\sigma}~.
\end{align}
The explicit construction is as follows:
\begin{subequations} \label{3.18}
\bea
& {\mathbb D}_\a := {\mathbb F}  \cD_\a - 2(\cD^\b {\mathbb F} ) \, M_{\a \b}  ~, \qquad
{\bar{\mathbb D}}_\ad := \bar{\mathbb F}  \bar \cD_\ad 
 -  2(\cDB^\bd \bar{\mathbb F} ) {\bar M}_{\bd\ad} ~, \\
& {\mathbb D}_{\a \ad} := \frac{\ri}{2} \big\{  {\mathbb D}_\a, {\bar{\mathbb D}}_\ad \big\}~.
\eea
\end{subequations} 
They obey the following (anti-)commutation relations:
\begin{subequations} \label{3.21}
\bea
\big\{ {\mathbb D}_\a , {\mathbb D}_\b \big\} &=& -4{\bar{\mathbb R}} M_{\a\b}~, \qquad 
\big\{ {\bar{\mathbb D}}_\ad , {\bar{\mathbb D}}_\bd \big\} = 4{\mathbb R} {\bar M}_{\ad\bd}~,\\
 \big[ {\bar{\mathbb D}}_\ad , {\mathbb D}_{\b \bd}\big] 
&=& \hf \ve_{\ad \bd} {\bar{\mathbb T}}^\gd {\mathbb D}_{\b \gd}
-\ri \ve_{\ad\bd} \Big( {\mathbb R} +\frac{1}{8} {\bar{\mathbb D}}_\gd {\bar{\mathbb T}}^\gd 
- \frac{1}{16} {\bar{\mathbb T}}_\gd {\bar{\mathbb T}}^\gd \Big) {{\mathbb D}}_\beta \non \\
&&-\ri \ve_{\ad\bd} \Big( {\mathbb G}_\b{}^\gd 
- \frac{1}{8} {\mathbb D}_\b {\bar{\mathbb T}}^\gd 
- \frac{1}{8} {\bar{\mathbb D}}^\gd {\mathbb T}_\b \Big) {\bar{\mathbb D}}_\gd
	- 2\ri \ve_{\dalpha \dbeta} {\mathbb W}_{\beta \gamma \delta} M^{\delta \gamma} +\cdots~~~~~~~
\eea
\end{subequations}
where the ellipsis denotes Lorentz curvature terms. The new torsion superfields are
constructed in terms of the old geometric quantities by
\begin{subequations} \label{3.22}
\bea
{\mathbb T}_\a &:=& {\mathbb D}_\a \ln (\mathbb F^4 \mathbb {\bar F}^2 ) ~, \label{nonmin-T}\\
{\mathbb R} &:=& -\frac{1}{4} (\bar \cD^2 - 4 R) \mathbb{\bar F}^2 ~,\label{nonmin-R}\\
{\mathbb W}_{\a \b \g} &:=& \mathbb {\bar F}^2 \mathbb F\,W_{\a\b\g}~,\\
{\mathbb G}_{\a \ad} &:=& \mathbb {F \bar F} \,G_{\a\ad} 
-\frac{1}{2} ({\mathbb D}_\a \ln { \mathbb {\bar F}}) {\bar{\mathbb D}}_\ad \ln {\mathbb F}
-\frac{1}{4} {\mathbb D}_\a  {\bar{\mathbb D}}_\ad \ln \frac{\mathbb {\bar F}^2}{\mathbb F}
+\frac{1}{4} {\bar{\mathbb D}}_\ad  {\mathbb D}_\a \ln \frac{\mathbb F^2}{\mathbb {\bar F}}~.~~~~~~~~~
\label{nonmin-G}
\eea
\end{subequations}
These new torsion superfields obey a number of Bianchi identities; we do not
reproduce them here, but refer the reader to \cite{BK}. The structure of torsion
superfields uncovered here is essentially that of conformal supergravity, where
the dilatations and $U(1)_R$ rotations have been ``degauged.'' (The Wess-Zumino
derivatives $\cD_A$ which we have used have an identical algebra but with
${\mathbb T}_\alpha = 0$.)

To describe non-minimal supergravity, we specialize to a specific compensator
$\mathbb F$ given in terms of a complex linear superfield $\Sigma$,
\begin{align}
{\mathbb F}= \big(\S^{n+1} \bar \S^{n-1}\big)^{-(3n+1)/8n}~.\label{3.19}
\end{align}
Using this relation along with the constraint \eqref{3.15}, one can check that an
additional relation on the torsion is imposed:
\bea
{\mathbb R} = -\frac{1}{4} \left(\frac{n+1}{3n+1}\right) {\bar{\mathbb D}}_\ad {\bar{\mathbb T}}^\ad 
+ \frac{1}{4} \Big(\frac{n+1}{3n+1} \Big)^2 {\bar{\mathbb T}}_\ad {\bar{\mathbb T}}^\ad ~.
\label{3.24}
\eea
This equation, we will see in the next section, is why an AdS solution cannot be consistently
constructed in non-minimal supergravity.

However, for the case $n=-1$ which concerns us, we shall make use of an improved
complex linear compensator, obeying the constraint \eqref{3.7}.
Replacing $\Sigma$ with $\Gamma$ in the above construction,
we find that the constraint \eqref{3.24} is replaced with
\bea
{\mathbb R} = W(\vf)~.
\label{3.23}
\eea
This will allow the construction of a non-minimal AdS supergravity, to which we now turn.

\section{Non-minimal AdS supergravity}
The novel formalism for non-minimal $n=-1$ supergravity, which we presented in 
the previous section, is well-suited to describe supergravity with a cosmological term
and its AdS solutions. This is achieved by demanding the conformal compensator
$\G$ to obey the  {\it super-Weyl invariant} constraint 
\bea
-\frac{1}{4} (\bar \cD^2 - 4 R) \Gamma &= \mu = {\rm const}
\label{4.1}
\eea
and choosing the following supergravity action 
\begin{align}
S_{\rm AdS} = -\fint (\Gamma \bar \Gamma)^{-1}~,
\label{4.2}
\end{align}
which is super-Weyl invariant.
Before discussing this theory, it is instructive to recall how $\cN=1$ AdS supergravity 
is described within the old minimal formulation for supergravity (see, e.g.,  \cite{BK} for more details).

AdS supergravity can be described using the super-Weyl invariant action
\begin{align}
S_{\rm AdS} = -3 \fint \phi \bar \phi + \mu \cint \phi^3
	+ \bar\mu \acint \bar\phi^3~,
\label{4.3}
\end{align}
where $\f$ is the chiral compensator with the super-Weyl transformation law (\ref{3.4}).
This corresponds to \eqref{3.1} for the choice $K=0$ and $W = \mu$.
The super-Weyl gauge freedom can always be used to fix 
\bea
\f =1~, 
\label{4.4}
\eea
and then (\ref{4.3}) 
 reduces to the standard action  for supergravity with cosmological term \cite{GGRS,BK}.
 
Let us analyse the equations of motion corresponding to the theory with action (\ref{4.3}).
The chirality constraint on $\phi$ and its equation of motion can
be written, respectively, as
\begin{subequations}\label{eq_phi}
\begin{align}
\bar \cD_\dalpha \phi^3 &= 0 \\
-\frac{1}{4} (\bar \cD^2 - 4 R) \left(\bar \phi \phi^{-2}\right) &= \mu~.
\label{4.5b}
\end{align}
\end{subequations}
We have written these equations in a peculiar way for reasons which
will soon be clear. In addition, there is an equation of motion from the
gravitational superfield\footnote{The gravitational superfield is the sole
physical prepotential for conformal supergravity.} $H^m$ \cite{OS,S}
\begin{align}
0 = G_{\alpha \dalpha} + (\phi \bar\phi)^{1/2} [\cD_\alpha, \bar\cD_\dalpha] (\phi \bar\phi)^{-1/2}~.
\label{4.6}
\end{align}
If we think of (\ref{4.3}) as a model for chiral matter $\f$
in a given supergravity background, then 
the equation (\ref{4.6}) is simply the condition that the matter supercurrent is equal
to zero. The equations (\ref{4.5b}) and (\ref{4.6}) are super-Weyl invariant. 
In the super-Weyl gauge (\ref{4.3}), they reduce to
\begin{align}
R = \mu~, \qquad G_{\alpha \dalpha} = 0~,
\label{4.7}
\end{align}
which are the equations of motion for $\cN=1$ AdS supergravity. 
AdS superspace is a conformally-flat solution of these equations,
with a vanishing superconformal Weyl tensor,
\bea
W_{\a\b\g}=0~.
\label{4.8}
\eea

Now let us turn to studying the equations of motion 
corresponding to the non-minimal supergravity action (\ref{4.2}).
The equation of motion for $\G$ and the constraint (\ref{4.1}) can be written
\begin{subequations}
\begin{align}
\bar \cD_\dalpha (\Gamma^2 \bar \Gamma)^{-1} &= 0 
\quad \Longleftrightarrow \quad \bar{\mathbb T}_\ad =0
 \label{4.9a} \\
-\frac{1}{4} (\bar \cD^2 - 4 R) \Gamma &= \mu~, \label{4.9b}
\end{align}
\end{subequations}
with $\bar{\mathbb T}_\ad $ the conjugate of (\ref{nonmin-T}). 
These equations are equivalent to \eqref{eq_phi} under the duality
$\phi^3 \leftrightarrow 1 / \Gamma^2 \bar \Gamma$. 
The equation of motion for the gravitational superfield can be shown
to be 
\bea
{\mathbb G}_{\a \ad} =0
\eea
with $ {\mathbb  G}_{\a\ad} $ the non-minimal curvature (\ref{nonmin-G}).
This equation is equivalent (using \eqref{4.9a}) to
\begin{align}
0 = G_{\alpha \dalpha} 
+ (\Gamma \bar\Gamma)^{-1/2} [\cD_\alpha, \bar\cD_\dalpha] (\Gamma \bar\Gamma)^{1/2}~.
\label{4.11}
\end{align}
It follows from (\ref{4.9a}) that 
\bea
\G = \f^{-2} \bar \f~, \qquad \bar \cD_\ad \f =0~
\eea
on the mass shell. Then, the super-Weyl  transformation law (\ref{3.10}) shows that 
we can impose the on-shell gauge condition 
\be
\G=1~.
\ee
In this gauge, the equations (\ref{4.9b}) and (\ref{4.11}) 
turn into the AdS equations (\ref{4.7}).
The AdS solution arises from the choice (\ref{4.8}).

It is now easy to explain why the traditional formulation of non-minimal supergravity 
\cite{GGRS,BK,DG} makes it impossible to generate the AdS geometry
\begin{subequations}\label{N=1-AdS-algebra}
\bea
&\{\cD_\a,\cD_\b\}=
-4\bar{\mu}M_{\a\b}~, \qquad \qquad
\{ \bar \cD_\ad , \bar \cD_\bd \} = 4\m \bar M_{\ad \bd}~,
\label{N=1-AdS-algebra-1}\\
&\{\cD_\a,{\bar \cD}_\bd\}=-2\ri(\s^{ c})_{\a \bd}\cD_{ c} \equiv -2\ri \cD_{\a\bd}
~,~~
\label{N=1-AdS-algebra-2}
\\
& {[}\cD_{ a},\cD_\b{]}=
-\frac{\ri}{ 2}   {\bar \mu} ({\s}_{ a})_{\b\gd} {\bar \cD}^\gd~,\qquad
{[}\cD_{ a}, \bar \cD_\bd{]}=
\frac{\ri}{ 2}   { \mu} ({\s}_{ a})_{\g \bd} { \cD}^\g~,
\label{N=1-AdS-algebra-3}
\\
& {[}\cD_{ a},\cD_{ b}{]}=-| { \mu}|^2 M_{ ab}~~~~~~~
\label{N=1-AdS-algebra-4}
\eea
\end{subequations}
as a globally (AdS) supersymmetric solution.
In order to derive (\ref{N=1-AdS-algebra}) from non-minimal supergeometry
associated with the covariant derivatives defined in (\ref{3.18}) with the choice
(\ref{3.19}), one has to keep the chiral torsion $\mathbb R$ and set to zero
the other components of the torsion, including ${\mathbb T}_\a$ and $\bar{\mathbb T}_\ad$.
But this is contradictory, since for a complex linear compensator $\mathbb R$ is
constructed in terms of $\bar{\mathbb T}_\ad$ in accordance with (\ref{3.24}).
If instead we use an improved complex linear compensator \eqref{4.1},
the torsion $\mathbb R$, eq. (\ref{nonmin-R}), is identically constant, 
${\mathbb R} = \m$, as a consequence of  (\ref{3.23}).

\section{Conformal supergravity and compensators}

In the preceding sections, we have used the Wess-Zumino superspace formulation
 involving the covariant derivative $\cD_A$ and torsion
superfields $W_{\alpha \beta \gamma}$, $R$ and $G_{\alpha \dalpha}$,
which describes the geometry of old minimal $(n=-1/3)$ supergravity.
This represents the most well-known in a family of different superspace
geometries, each well adapted to particular supergravity formulations.
Indeed, we have touched briefly upon this in section 3, where we
introduced the covariant derivative $\mathbb D_A$, with the additional
torsion superfield $\mathbb T_\alpha$, which is well-adapted to describe
non-minimal ($n\neq -1/3, 0$) supergravities.
In this section, we will elaborate upon this construction. We first
review the relationship between the various superspace formulations
and then discuss how the compensator method can be applied to conformal
supergravity to yield them.

\subsection{A brief survey of $\N=1$ superspace}
It was long ago realized at the component level that each of the
various supergravity formulations can be understood as conformal
supergravity, with $8+8$ degrees of freedom, coupled to various
compensators \cite{csg_poincare}.\footnote{See also the extensive
references in \cite{VanNieuwenhuizen:1981ae} and \cite{csg}.}
A chiral compensator, with $4+4$ degrees of freedom,
yields old minimal $(n=-1/3)$ supergravity with $12+12$ components; a real
linear compensator, also with $4+4$ components, yields new minimal
$(n=0)$ supergravity; and a complex linear compensator, with $12+12$
components, yields non-minimal $(n\neq -1/3, 0)$ supergravity with
$20+20$ components. Each of these supergravity formulations comes
with a corresponding supercurrent multiplet.

The Wess-Zumino superspace geometry, which we used
in section 3, describes old minimal supergravity with $12+12$ components.
It also economically describes conformal supergravity
at the component level. The super-Weyl parameter $\sigma$ contains
the component dilatation and $U(1)_R$ parameters along with the
spinor special conformal parameter (associated with the so-called
$S$-supersymmetry). As $\sigma$ possesses $4+4$ components, the
effective conformal degrees of freedom in the Wess-Zumino formulation
are precisely the $8+8$ components of conformal supergravity.
However, $\sigma$ is chiral and so does not represent the most
general super-Weyl transformation.

It is possible to enlarge the Wess-Zumino formulation to one which
allows an \emph{unconstrained} complex super-Weyl parameter. This
superspace, which was largely developed by Gates and Siegel \cite{SG,Siegel79,GS80}
and which we will refer to as conventional superspace,
is very similar to the Wess-Zumino formulation, but involves an
additional superfield $T_\alpha$. The algebra of covariant
derivatives is
\begin{subequations}\label{eq_SGalgebra}
\begin{align}
\{\CD_\beta, \CD_\alpha \} &= -4 {\bar R} \,M_{\beta \alpha}~, \qquad
\{\BCD^\dbeta, \BCD^\dalpha \}
	= 4 R \,\bar M^{\dbeta \dalpha}~, \qquad
\{\CD_\alpha, \BCD_\dalpha\} = -2i \CD_{\alpha \dalpha} \\
[\CD_\beta, \CD_{\alpha \dalpha}] &=
	\frac{1}{2} \eps_{\beta \alpha} T^\gamma \CD_{\gamma \dalpha}
	+ \ri \eps_{\beta \alpha}
		\left(\bar R + \frac{1}{8} \CD^\gamma T_\gamma - \frac{1}{16} T^\gamma T_\gamma\right)\BCD_\dalpha
	\eol & \quad
     - \ri \eps_{\beta \alpha} \left(G_{\gamma \dalpha} +
		\frac{1}{8} \CD_\gamma \bar T_\dalpha + \frac{1}{8} \BCD_\dalpha T_\gamma\right)\CD^\gamma
	+ \ri \BCD_\dalpha \bar R \,M_{\beta \alpha}
	\eol & \quad
     - \ri \eps_{\beta \alpha} \left(\CD_\delta - \frac{1}{2} T_\delta \right)
		\left(G_{\gamma \dalpha} + \frac{1}{8} \CD_\gamma \bar T_\dalpha + \frac{1}{8} \BCD_\dalpha T_\gamma\right)
		M^{\gamma \delta}
     \eol & \quad
     + 2\ri \eps_{\beta \alpha} \bar W_{\dalpha \dbeta \dgamma} \,\bar M^{\dbeta \gamma}
     - 2 \ri \eps_{\beta \alpha} \bar Y^\dbeta \bar M_{\dbeta \dalpha}
\end{align}
\end{subequations}
where $\bar Y^\dbeta$ is a particularly complicated set of derivatives
acting upon the superfields $T_\alpha$ and $\bar T_\dalpha$.\footnote{See e.g.
\cite{BK} for details. $\bar Y_\dalpha$ is denoted by $\bar X_\dalpha$ in \cite{BK}.}
This formulation of superspace naturally possesses $24+24$
degrees of freedom at the component level. However, it \emph{also}
may be understood to describe conformal supergravity at the component
level. This is because, like the Wess-Zumino formulation, it admits
a super-Weyl transformation; but in this case, it is an
\emph{unconstrained complex} super-Weyl transformation \cite{GSW} acting as
\begin{subequations}\label{eq_SGsw}
\begin{align}
\CD_\alpha' &= L \CD_\alpha - 2 \CD^\beta L M_{\beta \alpha} \label{eq_SGCDsw}\\
T'_\alpha &= L T_\alpha - \CD_\alpha \ln (L^4 \bar L^2) \label{eq_Tsw}\\
R' &= -\frac{1}{4} (\BCD^2 - 4 R) \bar L^2~ \label{eq_SGRsw}.
\end{align}
\end{subequations}
One may interpret 
the real part of $\log L$ 
as a real superfield parameter
associated with dilatations and 
the imaginary part of $\log L$ 
as a real superfield
parameter associated with chiral $U(1)_R$ rotations.
Since $L$ involves $16+16$ degrees of freedom, there remain $8+8$
conformal degrees of freedom corresponding to the Weyl multiplet
of conformal supergravity.

It is possible, due to eq. \eqref{eq_Tsw}, to choose $L$ such that
$T_\alpha$ vanishes. Doing so, the algebra above reduces to the
Wess-Zumino superspace algebra. One may make residual super-Weyl
transformations which preserve the gauge choice $T_\alpha=0$ provided
$L = \re^{\sigma / 2 - \bar\sigma}$ for chiral $\sigma$. This is exactly
the restricted chiral super-Weyl parameter of the Wess-Zumino superspace
formulation. This gauge choice describes old minimal supergravity.

It is also possible, by a suitable super-Weyl transformation, to
adopt a gauge where $R$ is defined in terms of $T_\alpha$:
\begin{align}
R = -\frac{1}{4} \left(\frac{n+1}{3n+1}\right) \BCD_\ad \bar T^\ad 
+ \frac{1}{4} \Big(\frac{n+1}{3n+1} \Big)^2 \bar T_\ad \bar T^\ad ~.
\end{align}
Because this choice fixes the components of $R$ in terms of
the components of $T_\alpha$, the theory has been reduced to
$20+20$ degrees of freedom. The resulting superspace structure
describes non-minimal supergravity.

However, this formulation does not easily describe new minimal
supergravity as it only explicitly gauges the Lorentz group and
new minimal supergravity involves the gauging of the $U(1)_R$ symmetry.
One requires $U(1)$ superspace introduced by Howe \cite{Howe}
(see \cite{GGRS} for a review).
 Its geometry is described by a
covariant derivative $\CD_A$, now carrying a $U(1)_R$ connection,
with an algebra
\begin{subequations}\label{eq_U1algebraS3}
\begin{align}
\{\CD_\beta, \CD_\alpha \} &= -4 {\bar R} \,M_{\beta \alpha}~, \qquad
\{\BCD^\dbeta, \BCD^\dalpha \}
	= 4 R \,\bar M^{\dbeta \dalpha}~, \qquad
\{\CD_\alpha, \BCD_\dalpha\} = -2 \ri \CD_{\alpha \dalpha} \\
[\CD_\beta, \CD_{\alpha \dalpha}] &=
     + \ri \eps_{\beta \alpha} \bar R \,\CD_\dalpha
     - \ri \eps_{\beta \alpha} G_{\gamma \dalpha} \,\CD^\gamma
     + \ri \bar\cD_\dalpha \bar R \,M_{\beta \alpha}
     - \ri \eps_{\beta \alpha} \cD_\delta G_{\gamma \dalpha} \,M^{\gamma \delta}
     \eol & \quad
     + 2\ri \eps_{\beta \alpha} \bar W_{\dalpha \dbeta \dgamma} \,\bar M^{\dbeta \gamma}
     - \frac{\ri}{3} \eps_{\beta \alpha} \bar X^\dbeta \bar M_{\dbeta \dalpha}
     + \frac{\ri}{2} \eps_{\beta \alpha} \bar X_\dalpha \,\A~.
\end{align}
\end{subequations}
The superfields $R$, $G_{\alpha \dalpha}$, $W_{\alpha \beta \gamma}$,
and $X_\alpha$ obey the relations
\begin{subequations}\label{eq_U1Torsion}
\begin{gather}
\BCD_\dalpha R = 0~, \qquad
\BCD_\dalpha X_\alpha = 0~, \qquad
\BCD_\dalpha W_{\alpha \beta \gamma} = 0 \\
X_\alpha = \CD_\alpha R - \BCD^\dalpha G_{\alpha \dalpha}~, \qquad
\CD^\alpha X_\alpha = \BCD_\dalpha \bar X^\dalpha~.
\end{gather}
\end{subequations}
The $U(1)_R$ generator $\A$ is normalized such that
\begin{align}
[\A, \CD_\alpha] = -\CD_\alpha~,\qquad [\A, \BCD_\dalpha] = +\BCD_\dalpha~.
\end{align}
The corresponding $U(1)_R$ transformations are gauged with all objects,
including the torsion superfields transforming with definite
$U(1)_R$ charge. In addition, one has an ungauged super-Weyl
transformation given by
\begin{align}
\CD_\alpha' &= \re^{\Lambda/2}
	\left(\CD_\alpha + 2 \CD^\beta \Lambda \,M_{\beta \alpha}
	+ \frac{3}{2} \CD_\alpha \Lambda \,\A\right)~.
\end{align}
The algebra \eqref{eq_U1algebraS3} and relations \eqref{eq_U1Torsion} are
much simpler than in the conventional formulation. The price one pays for
this simplification is that this superspace can only describe $U(1)_R$-invariant
theories.

The geometry naturally possesses $16+16$ degrees of freedom.
To describe new minimal supergravity, one can perform a super-Weyl
transformation to the gauge $R = 0$. This reduces the theory to
$12+12$ components and corresponds to new minimal supergravity.

It is possible to derive conventional superspace from $U(1)$ superspace.
One performs a ``degauging'' procedure where
the $U(1)_R$ connection is removed from the covariant derivative
and reinterpreted as a torsion superfield. This is, in a certain
sense, the origin of $T_\alpha$: it is ``really'' the spinor $U(1)_R$
connection $A_\alpha$  of $U(1)$ superspace. In order to maintain
the simple forms of the spinor anti-commutators in \eqref{eq_U1algebraS3},
one must redefine the superfield $R$ and the Lorentz connection (see, e.g.,
the discussion in \cite{GGRS}). These lead to the more complicated
algebra \eqref{eq_SGalgebra}. Moreover, after making these redefinitions,
one may show that the (now) ungauged $U(1)_R$ transformation combines
with the real super-Weyl transformation to yield the complex super-Weyl
transformation \eqref{eq_SGsw}.

In proceeding from the conventional superspace to $U(1)$ superspace,
we have gauged half of the original complex super-Weyl parameter.
It is natural to ask whether it is possible to gauge the remainder?
It turns out the answer is yes. However, instead of just enlarging the
structure group with dilatations, one must also gauge special \emph{superconformal}
transformations, involving not just the special conformal generator $K_a$
but also the $S$-supersymmetry generators $S_\alpha$ and $\bar S^\dalpha$.
The resulting superspace is called
conformal superspace \cite{Butter:2009cp}. Its geometry is characterized
by a covariant derivative $\nabla_A$ constructed out of connections valued
in the full superconformal algebra. It turns out that the algebra of the
covariant derivatives is quite simple, with all curvatures
constructed solely from the superconformal Weyl curvature
$W_{\alpha \beta \gamma}$. One finds\footnote{We remind the reader that we
make use of the conventions of \cite{BK} for
the Lorentz generator and normalization of $W_{\alpha \beta \gamma}$.
These conventions are different from those originally employed in
\cite{Butter:2009cp}.}
\begin{subequations}\label{eq_csgAlg}
\begin{gather}
\{\nabla_\alpha, \nabla_\beta\} = \{\bar\nabla_\dalpha, \bar\nabla_\dbeta\} = 0 ~,\qquad
\nabla_{\alpha \dalpha} := \frac{\ri}{2} \{\nabla_\alpha, \bar\nabla_\dalpha\} \label{eq_csgAlg1} \\
[\nabla_\beta, \nabla_{\alpha \dalpha}] =
	2i \eps_{\beta \alpha} \bar W_{\dalpha \dbeta \dgamma} \bar M^{\dbeta \dgamma}
	- R(\bar S)_{\beta\, \alpha \dalpha\,}{}_\dgamma \bar S^\dgamma
	- R(K)_{\beta\, \alpha \dalpha}{}^c K_c \label{eq_csgAlg2} 
\end{gather}
\end{subequations}
where $R(\bar S)_{\beta\, \alpha \dalpha\,}{}_\dgamma$ and $R(K)_{\beta\, \alpha \dalpha}{}^c$
involve derivatives of the superfield $\bar W_{\dalpha \dbeta \dgamma}$; their precise
forms will not be necessary for our discussion. The algebra of these
derivatives is considerably simplified while at the same time the
structure group has enlarged. All symmetries are gauged with
only the $8+8$ degrees of freedom of conformal supergravity surviving
at the component level. Of course, this huge simplification comes at
a high cost: only superconformally invariant actions can be described in
conformal superspace.

Just as conventional superspace may be derived by degauging $U(1)$ superspace,
the latter may be derived by degauging conformal superspace. Relative to $U(1)$
superspace, there are additional connections associated with
dilatations and the special superconformal transformations. The special
superconformal symmetry may be used to eliminate the dilatation connection
algebraically. The residual special superconformal connections can be
shown to yield precisely $R$, $G_{\alpha \dalpha}$ and $X_\alpha$ of
$U(1)$ superspace. The details are given in \cite{Butter:2009cp}. 

The degauging procedure which takes one from conformal superspace to $U(1)$ superspace
to conventional superspace is straightforward but messy. It requires
specific gauges to be imposed along the way and specific redefinitions
of connections to be made to keep the spinor anti-commutators simple.
A far more elegant method is available, which is similar to that used in
section 3 where we reviewed how to use an explicit compensator to convert
the Wess-Zumino formulation of superspace to conventional superspace (and
its non-minimal formulation in particular). We would like now to apply the same
idea to the construction of conventional superspace directly from conformal
superspace. It is natural to consider this procedure in two steps: deriving
$U(1)$ superspace first and then conventional superspace from $U(1)$.

\subsection{From conformal superspace to $U(1)$ superspace}
Let us assume that conformal superspace is furnished with a real,
nowhere-vanishing superfield $X = \exp U$ with non-vanishing dilatation
weight, which we assume to be $2$ without loss of generality. We
define new derivatives $\cUCD_A$ by\footnote{These new derivatives were
constructed originally by Kugo and Uehara at the level of tensor calculus
in their treatment of conformal supergravity \cite{Kugo2}. Their present
form appeared in \cite{Butter:2009wy}.}
\begin{subequations}\label{eq_U1D}
\begin{align}
\cUCD_\alpha &:= \nabla_\alpha - \frac{1}{2} \nabla_\alpha U \,\D
	- \nabla^\beta U \,M_{\beta \alpha}
	- \frac{3}{4} \nabla_\alpha U \, \A \\
\cUBCD_\dalpha &:= \bar\nabla_\dalpha - \frac{1}{2} \bar\nabla_\dalpha U \,\D
	- \bar\nabla^\dbeta U \,\bar M_{\dbeta \dalpha}
	+ \frac{3}{4} \bar\nabla_\dalpha U \,\A \\
\cUCD_{\alpha \dalpha} &:= \frac{\ri}{2} \{\cUCD_\alpha, \cUBCD_\dalpha\}~,
\end{align}
\end{subequations}
where $\D$ and $\A$ are the dilatation and $U(1)_R$ generators,
normalized such that
\begin{subequations}
\begin{align}
[\D, \nabla_\alpha] = \frac{1}{2} \nabla_\alpha~,\qquad
[\D, \bar\nabla_\dalpha] = \frac{1}{2} \bar\nabla_\dalpha \\
[\A, \nabla_\alpha] = -\nabla_\alpha~,\qquad
[\A, \bar\nabla_\dalpha] = +\bar\nabla_\dalpha~.
\end{align}
\end{subequations}
The new derivatives $\cUCD_A$ have the same dilatation and $U(1)_R$ weights
as $\nabla_A$. However, \emph{unlike} $\nabla_A$, they have the property that
they act on conformally primary superfields to give new conformally primary
superfields. Recall that a superfield $\Psi$ is called \emph{primary} if it
is annihilated by the special superconformal generators,
\begin{align}
K_B \Psi = 0~, \qquad K_B = (K_b, S_\beta, \bar S^\dbeta)~.
\end{align}
One can show that the derivative of $\Psi$, constructed using
$\cUCD_A$, is \emph{also} primary,
\begin{align}\label{eq_KFtoKDF}
K_B \Psi= 0 \implies K_B \cUCD_A \Psi = 0~.
\end{align}
This requirement uniquely determines the definition of $\cUCD_\alpha$ and
$\cUBCD_\dalpha$ in \eqref{eq_U1D}.
As we will show, they are closely related to the
super-Weyl invariant derivatives constructed in
\cite{BKS}.\footnote{See also the discussion in \cite{BK} for a pedagogical
review.}
Note that if we are careful to only work with conformally primary superfields
$\Psi$ and their derivatives $\cUCD_A \Psi$, then the special superconformal
generators appearing in $\cUCD_A$ are purely spectator connections. In particular,
when we discuss the curvatures $[\cUCD_A, \cUCD_B]$, we will always assume
that we are acting on conformally primary superfields and so we can always
ignore the special superconformal curvatures.

There is another key feature of these new derivatives. The compensator $X$
is covariantly constant with respect to them,
\begin{align}
\cUCD_\alpha X = \cUBCD_\dalpha X = 0~.
\end{align}
This implies that
\begin{align}
[\cUCD_A, \cUCD_B] X = 0
\end{align}
and so the only curvatures present in $[\cUCD_A, \cUCD_B]$ must be those for which $X$
is a scalar. Thus we are left with torsion, Lorentz, and $U(1)_R$ curvatures;
there can be no dilatation curvatures. In fact, one can show that when these
new derivatives act \emph{only on conformally primary objects}, they obey the same
algebra as \eqref{eq_U1algebraS3}
\begin{subequations}\label{eq_U1algebra}
\begin{align}
\{\cUCD_\beta, \cUCD_\alpha \} &= -4 {\bar \cUR} \,M_{\beta \alpha}~, \qquad
\{\cUBCD^\dbeta, \cUBCD^\dalpha \}
	= 4 \cUR \,\bar M^{\dbeta \dalpha} \\
[\cUCD_\beta, \cUCD_{\alpha \dalpha}] &=
     - \ri \eps_{\beta \alpha} \cUG_{\gamma \dalpha} \,\cUCD^\gamma
     + \ri \eps_{\beta \alpha} \bar \cUR \,\cUCD_\dalpha
     - \ri \eps_{\beta \alpha} \cUCD_\delta \cUG_{\gamma \dalpha} \,M^{\gamma \delta}
     + \ri \cUBCD_\dalpha \bar \cUR \,M_{\beta \alpha}
     \eol & \quad
     + 2\ri \eps_{\beta \alpha} \bar W_{\dalpha \dbeta \dgamma} \,\bar M^{\dbeta \gamma}
     - \frac{\ri}{3} \eps_{\beta \alpha} \bar \cUX^\dbeta \bar M_{\dbeta \dalpha}
     + \frac{\ri}{2} \eps_{\beta \alpha} \bar \cUX_\dalpha \,\A~.
\end{align}
\end{subequations}
where we have defined
\begin{align}
\cUR := -\frac{1}{4X} \bar \nabla^2 X~, \qquad
\cUG_{\alpha \dalpha} := X^{1/2} [\nabla_\alpha, \bar\nabla_\dalpha] X^{-1/2}, \qquad
\cUX_\alpha := \frac{3}{4} \bar\nabla^2 \nabla_\alpha \log X~.
\end{align}
The superfield $W_{\alpha \beta \gamma}$ is identical to that appearing in
the original conformal superspace algebra \eqref{eq_csgAlg}.
Each of the new torsion superfields is conformally primary.
They can be shown to obey the constraints
\begin{subequations}
\begin{align}
\cUBCD_\dalpha \cUR = 0~, \qquad
\cUBCD_\dalpha \cUX_\alpha = 0~, \qquad
\cUBCD_\dalpha W_{\alpha \beta \gamma} = 0 \\
\cUX_\alpha = \cUCD_\alpha \cUR - \cUBCD^\dalpha \cUG_{\alpha \dalpha}~, \qquad
\cUCD^\alpha \cUX_\alpha = \BCD_\dalpha \bar \cUX^\dalpha
\end{align}
\end{subequations}
which are identical in form to \eqref{eq_U1Torsion}.
This structure is exactly that of $U(1)$ superspace.

One may make a further set of redefinitions which clarifies that
dilatations have essentially been removed from the structure group along with
special conformal transformations. For every superfield $\Psi$ (including our
torsion superfields) with dilatation weight $\Delta$, we may define a new
superfield $\UPsi$ by
\begin{align}
\UPsi := \re^{-U \D /2} \cF = X^{-\Delta / 2} \Psi
\end{align}
which is inert under superfield dilatations. For our torsion superfields,
this amounts to
\begin{subequations}\label{eq_UTorsion}
\begin{align}
\UR := X^{-1/2} \cUR = -\frac{1}{4X^{3/2}} \bar \nabla^2 X~, \qquad
\UG_{\alpha \dalpha} := X^{-1/2} \cUG_{\alpha \dalpha} = [\nabla_\alpha, \bar\nabla_\dalpha] X^{-1/2}\\
\UX_\alpha := X^{-3/4} \cUX_\alpha = \frac{3}{4} X^{-3/4} \bar\nabla^2\nabla_\alpha \log X~, \qquad
\UW_{\alpha \beta \gamma} := X^{-3/4} W_{\alpha \beta \gamma}~.
\end{align}
\end{subequations}
Similarly, the operator
$\cUCD_\alpha$, which carries dilatation weight $1/2$, may be used to define a
new operator $\UCD_\alpha$ via the similarity transformation
\begin{align}\label{eq_DCompD}
\UCD_\alpha := \re^{-U \D/2} \,\cUCD_\alpha\, \re^{U \D/2}
	= X^{-1/4} \left(\nabla_\alpha 
	- \nabla^\beta U M_{\beta \alpha}
	- \frac{3}{4} \nabla_\alpha U \,\A \right)~.
\end{align}
These new derivatives \emph{also} obey the $U(1)$ superspace algebra,
but with all the objects modified by the similarity transformation.

These new objects are inert under the dilatations of conformal
superspace; we say that the superfield $X$ has compensated for
the dilatations. However, there is a new transformation that
the compensated objects possess. If we consider a finite shift
in the compensator $X$
\begin{align}
X \rightarrow \re^{-2 \Lambda} X
\end{align}
the new covariant derivative transforms as
\begin{align}
\UCD_\alpha & \rightarrow \re^{\Lambda/2}
	\left(\UCD_\alpha + 2 \UCD^\beta \Lambda \,M_{\beta \alpha}
	+ \frac{3}{2} \UCD_\alpha \Lambda \,\A\right)~.
\end{align}
This is exactly the super-Weyl transformation of $U(1)$ superspace.
It manifests itself here from an explicit shift in the compensator.

This additional similarity transformation is quite natural since
it absorbs the compensator into the frame of superspace by the redefinition
of the vierbein implicit in \eqref{eq_DCompD}. For example, given a
superconformal action
\begin{align}\label{eq_S1}
S = \fint X^{1 - \Delta/2} \cL
\end{align}
where $\cL$ is some quantity with dilatation weight $\Delta$, the derivatives
$\cUCD_A$ treat $X$ as if it were constant. Under the similarity
transformation, the action becomes
\begin{align}\label{eq_S2}
S = \int \rd^4x\, \rd^4\theta\, {\bm E}\, \UL
\end{align}
with all explicit factors of $X$ being absorbed either into other
superfields or the frame of superspace. This agrees with the general notion
of a compensator field as discussed in \cite{GGRS} where a compensator
field is explicitly used to construct the vierbein.

The geometry of $U(1)$ superspace is well-adapted to describe new minimal
supergravity constructed with a linear compensator. However,
since our concern is with old minimal and non-minimal geometry, we will
proceed to degauge to conventional superspace.

\subsection{From $U(1)$ superspace to conventional superspace}
Now we turn to the task of degauging the $U(1)_R$ connection.
The simplest way of achieving this is again via the use of an explicit compensator field.
In analogy to the real compensator $X$ which transforms under dilatations with some weight,
we could introduce a real compensator, say $Y$, transforming under
$U(1)_R$ chiral rotations with some weight.
However, describing the geometry in terms of two separate real compensators
is somewhat artificial. It is more natural to consider a single complex superfield
$\mathbb F$ with non-vanishing dilatation and $U(1)_R$ weights. Without loss of generality,
we take those to be $-1/2$ and $1$, so that the dilatation compensator $X$ used
in $U(1)$ superspace is given by
\begin{align}
X = (\mathbb F \bar {\mathbb F})^{-2}~.
\end{align}
The reason for this choice is so that subsequent formulae simplify.

It turns out that the presence of both dilatation and $U(1)_R$ compensators
offers a new dimension-1/2 conformally primary superfield in addition to the
superfields defined previously.
This combination, which we denote $\cFT_\alpha$, can be written
in terms of the $U(1)$ superspace spinor derivative $\cUCD_\alpha$ as
\begin{align}
\cFT_\alpha := \cUCD_\alpha \log (\mathbb F / \bar {\mathbb F})~.
\end{align}
Equivalently, we can write this quantity using the original conformal superspace
derivative as
\begin{align}
\cFT_\alpha = \nabla_\alpha \log (\mathbb F^4 \bar{\mathbb F}^2)~.
\end{align}
It is now possible to degauge the $U(1)_R$ symmetry by introducing
the new covariant derivatives
\begin{subequations}
\begin{align}
\cFCD_\alpha &:= \cUCD_\alpha - \cFT^\beta M_{\beta \alpha}
	- \frac{1}{2} \cFT_\alpha \A \\
\cFBCD_\dalpha &:= \cUCD_\dalpha
	- \bar \cFT^\dbeta \bar M_{\dbeta \dalpha}
	+ \frac{1}{2} \bar \cFT_\dalpha \A \\
\cFCD_{\alpha \dalpha} &:= \frac{\ri}{2} \{\cFCD_\alpha, \bar \cFCD_\dalpha\}~.
\end{align}
\end{subequations}
The modification of the $U(1)_R$ connection is necessary so that the new derivatives obey
\begin{align}
\cFCD_A \mathbb F = 0
\end{align}
which guarantees that no $U(1)_R$ curvatures will appear in the (anti-)commutator
$[\cFCD_A, \cFCD_B]$. Unlike in the previous section, there is a certain degree of
arbitrariness in the definitions made for $\cFCD$ -- specifically, in the choice of
how to deform the Lorentz connection. The reason for this is that $\cFT_\alpha$ is already
conformally primary, so there is no unambiguous choice for $\cFCD_A$. The coefficient
we have chosen is so that the anti-commutator of $\cFCD_\alpha$ with itself
is simply
\begin{align}
\{\cFCD_\alpha, \cFCD_\beta\}
	= - 4 \bar \cFR M_{\beta \alpha}~,\qquad
\{\cFBCD_\dalpha, \cFBCD_\dbeta\}
	= + 4 \cFR \bar M_{\dbeta \dalpha}~.
\end{align}
where
\begin{align}
\cFR := \cUR 
	- \frac{1}{4} \cUBCD_\dalpha \bar\cFT^\dalpha
	- \frac{1}{4} \bar\cFT_\dalpha \bar\cFT^\dalpha~.
\end{align}
Any other choice would introduce dimension-1/2 torsion. In \cite{GGRS} this is understood
as a certain arbitrariness in how one chooses to define the spin connection, or
equivalently, an arbitrariness in what specific constraints to place upon the
torsion. For the remainder of the (anti-)commutators, we find a result identical
to that given in \eqref{3.21} and \eqref{eq_SGalgebra}
provided we make the following definition for the superfield $\cFG_{\alpha \dalpha}$:
\begin{align}
\cFG_{\alpha \dalpha} &:= \cUG_{\alpha \dalpha}
	- \frac{3}{8} \cUCD_\alpha \bar \cFT_\dalpha
	+ \frac{3}{8} \cUBCD_\dalpha \cFT_\alpha
	+ \frac{1}{4} \cFT_\alpha \bar \cFT_\dalpha~.
\end{align}
This quantity may be expressed in terms of conformal superspace derivatives
and the explicit compensators as
\begin{align}
\cFG_{\alpha \dalpha}
	&= -\frac{1}{4} \nabla_\alpha \bar\nabla_\dalpha \log\frac{\bar {\mathbb F}^2}{\mathbb F}
	+ \frac{1}{4} \bar\nabla_\dalpha \nabla_\alpha \log\frac{\mathbb F^2}{\bar {\mathbb F}}
	\eol & \quad
	- \frac{1}{4} \nabla_\alpha \log({\mathbb F} \bar {\mathbb F})
		\bar\nabla_\dalpha \log({\mathbb F} \bar {\mathbb F})
	- \frac{1}{4} \nabla_\alpha \log({\mathbb F} /  \bar {\mathbb F})
		\bar\nabla_\dalpha \log({\mathbb F} / \bar {\mathbb F})
\end{align}
In a similar vein, the superfield $\cFR$ may be written as
\begin{align}
\cFR &= -\frac{1}{4} \bar {\mathbb F}^{-2} \bar\nabla^2 \bar {\mathbb F}^2~.
\end{align}

As in our construction of $U(1)_R$ superspace, these new derivatives and
superfields are not \emph{exactly} those of conventional superspace. One
must perform an additional similarity transformation involving both the
dilatation and $U(1)_R$ compensators on all operators and superfields to
remove the dilatation and $U(1)_R$ connections completely. For
the covariant derivative, we find
\begin{align}
\FCD_\alpha &= \mathbb F \nabla_\alpha
	- 2 \nabla^\beta \mathbb F M_{\beta \alpha}
\end{align}
after the appropriate similarity transformation. The torsion superfields are
given by
\begin{subequations}
\begin{align}
\FT_\alpha &= \mathbb F \,\nabla_\alpha \log (\mathbb F^4 \bar{\mathbb F}^2) \\
\FR &= -\frac{1}{4} \bar\nabla^2 \bar {\mathbb F}^2 \\
\FG_{\alpha \dalpha} &= 
	-\frac{1}{4} \mathbb {F \bar F} \nabla_\alpha \bar\nabla_\dalpha \log\frac{\bar {\mathbb F}^2}{\mathbb F}
	+ \frac{1}{4} \mathbb {F \bar F} \bar\nabla_\dalpha \nabla_\alpha \log\frac{\mathbb F^2}{\bar {\mathbb F}}
	\eol & \quad\quad\quad
	- \frac{1}{4} \mathbb {F \bar F} \nabla_\alpha \log({\mathbb F} \bar {\mathbb F})
		\bar\nabla_\dalpha \log({\mathbb F} \bar {\mathbb F})
	- \frac{1}{4} \mathbb {F \bar F} \nabla_\alpha \log({\mathbb F} /  \bar {\mathbb F})
		\bar\nabla_\dalpha \log({\mathbb F} / \bar {\mathbb F}) \\
\FW_{\alpha \beta \gamma} &= \bar {\mathbb F}^2 \mathbb F \,W_{\alpha \beta \gamma}~.
\end{align}
\end{subequations}
If we consider arbitrary redefinitions of the compensator,
\begin{align}
\mathbb F \rightarrow L \mathbb F
\end{align}
we find the complex super-Weyl transformations
\begin{subequations}
\begin{align}
\FCD_\alpha &\rightarrow L \FCD_\alpha - 2 \FCD^\beta L M_{\beta \alpha} \\
\FT_\alpha &\rightarrow L \FT_\alpha + \FCD_\alpha \log (L^4 \bar L^2) \\
\FR &\rightarrow -\frac{1}{4} (\FBCD^2 - 4 \FR) \bar L^2 ~.
\end{align}
\end{subequations}
In this formulation, all operators and
superfields have been rendered completely inert under both dilatations \emph{and}
the $U(1)_R$ rotations. In their place we are left with a complex super-Weyl
transformation arising from shifts in the complex compensator $\mathbb F$.

\subsection{Old minimal, non-minimal and improved non-minimal supergravity}
There are two immediate choices of complex compensator which can be made. The
most obvious choice is a chiral compensator. If we choose $\mathbb F = \phi^{1/2} \bar\phi^{-1}$,
we immediately observe that
\begin{align}
\FT_\alpha = \bar\FT_\dalpha = 0~.
\end{align}
The geometry of Wess and Zumino which we have used in sections 2 through 4
corresponds to such a conventional superspace description.

A more interesting possibility is offered by the family of non-minimal
supergravities. These coincide with the choice 
\begin{align}
{\mathbb F}= \big(\S^{n+1} \bar \S^{n-1}\big)^{-(3n+1)/8n}
\end{align}
for a complex linear multiplet. Recall that the dilatation and $U(1)_R$ weights
of $\Sigma$ are parametrized by a single real number $n$,
\begin{align}\label{eq_CLweights2}
\Delta = \frac{2}{3n + 1}~, \quad w = -\frac{4n}{3n + 1}~, \qquad n \neq -1/3,0~.
\end{align}
Now, because of the constraint $\bar\nabla^2 \Sigma = 0$
one can show that the superfield $\FR$ is actually given in
terms of $\bar\FT_\dalpha$ as
\begin{align}\label{eq_nonminR}
\FR = -\frac{1}{4} \left(\frac{n+1}{3n+1}\right) \FBCD_\dalpha
	\bar\FT^\dalpha
	+ \frac{1}{4} \left(\frac{n+1}{3n+1}\right)^2 \bar\FT_\dalpha \bar\FT^\dalpha~.
\end{align}
This relation defines the so-called non-minimal superspaces.

Finally, we consider the case where the superfield $\Gamma$ obeys the
improved complex linearity constraint
\begin{align}
-\frac{1}{4} \bar\nabla^2 \Gamma = Q
\end{align}
where $Q$ is some chiral superfield. We will be most interested in the case where it
is a holomorphic function of some chiral matter superfields, but for the moment we
will allow it to be arbitrary. $\Gamma$ has dilatation and $U(1)_R$ weights parametrized
by a real number $n$ \eqref{eq_CLweights2} and so $Q$ must have dilatation and $U(1)_R$ weights given by
\begin{align}
\Delta(Q) = 3 \left(\frac{n+1}{3n+1}\right)~, \qquad
w(Q) = 2 \left(\frac{n+1}{3n+1}\right)~.
\end{align}
Note that $\Delta(Q)$ and $w(Q)$ are in the ratio of $3 : 2$ as required for
$Q$ to be chiral. One can show that in this situation, the
constraint \eqref{eq_nonminR} is modified to
\begin{align}\label{eq_defnonminR}
\FR = \mathbb Q
	-\frac{1}{4} \left(\frac{n+1}{3n+1}\right) \FBCD_\dalpha \bar\FT^\dalpha
	+ \frac{1}{4} \left(\frac{n+1}{3n+1}\right)^2 \bar\FT_\dalpha \bar\FT^\dalpha~,
\end{align}
where $\mathbb Q$ is the superfield $Q$ dressed with the appropriate
factors of the compensator $\mathbb F$ to render it inert,
\begin{align}
\mathbb Q:= (\mathbb F \bar {\mathbb F})^{\Delta(Q)} \left(\frac{\bar {\mathbb F}}{\mathbb F}\right)^{w(Q)/2} Q~.
\end{align}

\section{The geometry of matter couplings in non-minimal supergravity}
We turn now to our final topic: the geometrization of the dual action
\eqref{3.14}, which for convenience we repeat here
\begin{align}
S_{\rm dual} = -\int \rd^4x\, \rd^4\theta\, E\,{\rm e}^{-K} (\Gamma \bar\Gamma)^{-1}~.
\end{align}
The superfield $\Gamma$ obeys the constraint
\begin{align}
-\frac{1}{4} (\bar\cD^2 - 4 R) \Gamma = W
\end{align}
when written in terms of Wess-Zumino derivatives. Using conformal
superspace derivatives, this constraint is simply
\begin{align}
-\frac{1}{4} \bar\nabla^2 \Gamma = W~.
\end{align}

Recall that this action arose by performing a duality transformation on
the action
\begin{align}
S = -3 \fint \phi \bar\phi \,\re^{-K/3}
	+ \cint \phi^3 W
	+ \acint \bar\phi^3 \bar W~.
\end{align}
There exists a quite elegant formulation for dealing with this model,
known as K\"ahler superspace, which was constructed originally in
\cite{Binetruy:1987qw}.\footnote{See \cite{Binetruy:2000zx} for a 
recent detailed and pedagogical discussion.} As a first step toward
understanding how to geometrize the dual action, we will describe how
K\"ahler superspace arises via the compensator method.

The first step in this procedure is to identify the function $K$ as a
(composite) prepotential for the symmetry $U(1)_K$. The superfields
$W$ and $\phi$ are naturally understood as sections transforming under this symmetry
and are \emph{conventionally} chiral with respect to it. We may introduce \emph{covariantly}
chiral superfields $\Phi$ and $\cW$, which in the Hermitian basis are defined by
\begin{align}
\Phi := \re^{-K / 6} \phi ~, \qquad \cW := \re^{K/2} W \qquad \qquad \textrm{(Hermitian basis)}~.
\end{align}
This requires that we introduce a $U(1)_K$ connection into our covariant derivative;
for the conformal superspace derivative, we have
\begin{align}
\nabla_A' = \nabla_A - \ri A_A^{(K)} \K
\end{align}
(henceforth dropping the prime) where the $U(1)_K$ generator $\K$ acts on $\Phi$ and $\cW$ as
\begin{align}
\K \Phi = -\frac{2}{3}\Phi~, \qquad 
\K \cW = 2 \cW~.
\end{align}
In the Hermitian basis, the new connection is given by
\begin{align}
A_\alpha^{(K)} = \frac{\ri}{4} \nabla_\alpha K~,\quad
\bar A_\dalpha^{(K)} = -\frac{\ri}{4} \bar\nabla_\dalpha K~,\quad
A_{\alpha \dalpha}^{(K)} = \frac{1}{8} [\nabla_\alpha, \bar\nabla_\dalpha] K
\quad \textrm{(Hermitian basis)}
\end{align}
In \emph{any} basis, the action becomes
\begin{align}\label{eq_Ksuper}
S = -3 \fint \Phi \bar\Phi
	+ \cint \Phi^3 \cW
	+ \acint \bar\Phi^3 \bar \cW
\end{align}
where the K\"ahler potential is now implicit in the chirality constraint
on $\Phi$
\begin{align}
\bar\nabla_\dalpha \Phi = 0~.
\end{align}
The covariant derivatives of conformal superspace, augmented with
a $U(1)_K$ connection, now obey the constraints \eqref{eq_csgAlg1}
with \eqref{eq_csgAlg2} modified to read
\begin{align}
[\nabla_\beta, \nabla_{\alpha \dalpha}] = 
	\frac{\ri}{2} \eps_{\beta \alpha} \bar K_\dalpha \K
	+ \ldots~, \qquad \bar K_\dalpha := -\frac{1}{4} \nabla^2 \bar\nabla_\dalpha K~,
\end{align}
where the ellipsis denotes the previous set of terms in \eqref{eq_csgAlg2}.

We now proceed to apply the compensator procedure as outlined in section 5.
However, there is one crucial difference. Before, the complex compensator $\mathbb F$
transformed only under dilatations and $U(1)_R$ transformations.
Since here we wish to take $\mathbb F = \Phi^{1/2} \bar\Phi^{-1}$,
it will transform under dilatations, $U(1)_R$, \emph{and}
$U(1)_K$ transformations. By examining the $U(1)_K$ and $U(1)_R$
charges of $\Phi$, we see that 
\begin{align}
(\K + \A) \Phi = 0 \implies (\K + \A) \mathbb F = 0~.
\end{align}
Thus there is no reason why curvatures involving the \emph{combination}
$\K + \A$ cannot appear in $[\mathbb D_A, \mathbb D_B]$.
In fact, we find that they do.\footnote{In particle physics parlance,
$\Phi$ has ``higgsed'' the group $U(1)_{R} \times U(1)_K$ to the unbroken
subgroup $U(1)_{R+K}$.} An explicit calculation yields
\begin{align}
[\FCD_\beta, \FCD_{\alpha \dalpha}] =
	\frac{\ri}{2} \eps_{\beta \alpha} \bar \FK_\dalpha\,\left(\A+\K\right)
	+ \ldots
\end{align}
where the ellipsis refers to terms we had before. We still have
\begin{align}
\FT_\alpha = \mathbb F \nabla_\alpha \log(\mathbb F^4 \bar{\mathbb F}^2) =
	-3 \,\mathbb F \,\nabla_\alpha \log \bar\Phi = 0
\end{align}
as in Wess-Zumino superspace. It turns out that the algebra is (nearly) identical
to the algebra of $U(1)$ superspace, but with the generator $\A$ replaced
by $\A + \K$ and the curvature superfield $\UX_\alpha$ identified with $\FK_\alpha$,
reflecting the fact that only the group $U(1)_{R + K}$ is gauged.
In fact, the generator $\A$ may even be dropped from the combination $\A + \K$
since there are no longer any objects in the theory with $U(1)_R$ weight;
they have all been modified with appropriate factors of the complex compensator
to be $U(1)_R$ inert. Essentially, the complex compensator has transmuted
all $U(1)_R$-charged objects into corresponding $U(1)_K$-charged ones.

Armed with this insight, we may consider now the geometry associated with
a complex linear multiplet carrying K\"ahler weight. This appears naturally
if we consider the dual of the model \eqref{eq_Ksuper}. In first order form, we introduce
the action
\begin{align}
S = \fint \Big(-3 \Phi \bar\Phi
	+ \Gamma \Phi^3 + \bar\Gamma \bar\Phi^3\Big)
\end{align}
where $\Phi$ and $\Gamma$ transform \emph{covariantly} under both $U(1)_K$ and $U(1)_R$.
$\Phi$ is otherwise an unconstrained complex superfield while $\Gamma$ obeys
\begin{align}\label{eq_KGammaConstraint}
-\frac{1}{4} \bar\nabla^2 \Gamma = \cW
\end{align}
where $\cW$ is the K\"ahler-covariant superpotential. The equation of motion
for $\Gamma$ enforces that $\Phi$ is covariantly chiral. Integrating
out $\Phi$ instead using
\begin{align}
\Phi^3 = \Gamma^{-2} \bar\Gamma^{-1}
\end{align}
gives the dual action
\begin{align}
S_{\rm dual} = -\fint (\Gamma \bar\Gamma)^{-1}
\end{align}
with $\Gamma$ obeying the constraint \eqref{eq_KGammaConstraint}.

If we now choose the complex compensator $\mathbb F = \bar\Gamma^{1/2}$ we
find the algebra
\begin{subequations}
\begin{align}
\{\FCD_\beta, \FCD_\alpha \} &= -4 {\bar \FR} \,M_{\beta \alpha}~, \qquad
\{\FBCD^\dbeta, \FBCD^\dalpha \}
	= 4 \FR \,\bar M^{\dbeta \dalpha}~, \qquad
\{\FCD_\alpha, \FBCD_\dalpha\} = -2\ri \FCD_{\alpha \dalpha} \\
[\FCD_\beta, \FCD_{\alpha \dalpha}] &=
	\frac{1}{2} \eps_{\beta \alpha} \FT^\gamma \FCD_{\gamma \dalpha}
	+ \ri \eps_{\beta \alpha}
		\left(\bar \FR + \frac{1}{8} \FCD^\gamma \FT_\gamma - \frac{1}{16} \FT^\gamma \FT_\gamma\right)\FBCD_\dalpha
	\eol & \quad
     - \ri \eps_{\beta \alpha} \left(\FG_{\gamma \dalpha} +
		\frac{1}{8} \FCD_\gamma \bar \FT_\dalpha + \frac{1}{8} \FBCD_\dalpha \FT_\gamma\right)\FCD^\gamma
	+ \ri \FBCD_\dalpha \bar \FR \,M_{\beta \alpha}
	\eol & \quad
     - \ri \eps_{\beta \alpha} \left(\FCD_\delta - \frac{1}{2} \FT_\delta \right)
		\left(\FG_{\gamma \dalpha} + \frac{1}{8} \FCD_\gamma \bar \FT_\dalpha + \frac{1}{8} \FBCD_\dalpha \FT_\gamma\right)
		M^{\gamma \delta}
     \eol & \quad
     + 2\ri \eps_{\beta \alpha} \bar \FW_{\dalpha \dbeta \dgamma} \,\bar M^{\dbeta \gamma}
     - 2 \ri \eps_{\beta \alpha} \bar {\mathbb Y}^\dbeta \bar M_{\dbeta \dalpha}
	+ \frac{\ri}{2} \eps_{\beta \alpha} \bar \FK_\dalpha\,\left(\A+\K\right)
\end{align}
\end{subequations}
Because all objects carrying $U(1)_R$ charge have been modified with
factors of the compensator so that they now carry only $U(1)_K$ charge, we
may dispense with the generator $\A$ in the above algebra if we like.
Making use of \eqref{eq_defnonminR}, we find the constraint
\begin{align}
\FR = \cW~.
\end{align}
The K\"ahler potential and superpotential have been completely absorbed
into the geometry of superspace. In addition we have the simple action
\begin{align}
S_{\rm dual} = -\int \rd^4x\, \rd^4\theta\, \mathbb E~,
\end{align}
containing non-minimal supergravity, the matter kinetic terms, \emph{and}
the matter superpotential.

This formulation can be naturally understood as the dual of the original
K\"ahler superspace.  There one has the geometric constraint $\FT_\alpha = 0$
and (one can show \cite{Binetruy:2000zx}) the compensator equation of motion
$\FR = \cW$. In the dual formulation we have just constructed, we have the reverse: $\FR = \cW$ is a
constraint while (one can show) $\FT_\alpha = 0$ becomes the compensator
equation of motion. This is exactly what is expected in a dual formulation.

\section{Concluding remarks and AdS supercurrents}
The dual formulation we have constructed is quite interesting, if only because
it resolves the long-standing puzzle of finding an AdS formulation of non-minimal
supergravity which linearizes to the transverse superspin-3/2 model in AdS, eq. (\ref{eq_transAdS}).
We have also shown one way to see why only the $n=-1$ version of non-minimal
supergravity has such a linearized action: it is the only version of supergravity
for which we can deform the constraint $(\bar\cD^2 -4R) \Sigma = 0$ to
$(\bar\cD^2 -4R)\Gamma = -4 \mu = \textrm{const}$ in a super-Weyl invariant
way.\footnote{For a general $n\neq -1$, we can consider a deformation $(\BCD^2 - 4 R) \Gamma = -4 Q$ where
$Q$ is a chiral scalar with nontrivial super-Weyl transformation.}

More importantly, however, it emphasizes a key feature of AdS which we discussed
in \cite{BK2011}: there are two irreducible supercurrents, and each one corresponds
to some formulation of supergravity in AdS, either old minimal or non-minimal.
The supercurrent associated with old minimal supergravity is the Ferrara-Zumino
multiplet \cite{FZ}, obeying
\begin{align}\label{eq_oldminCurrent}
\BCD^\dalpha J_{\alpha \dalpha} = \CD_\alpha X~, \qquad \BCD_\dalpha X = 0~,
\end{align}
where $J_{\alpha \dalpha}$ is real and $X$ is chiral. These superfields may be
calculated in any model by varying the action with respect
to the prepotential $H_{\alpha \dalpha}$ and the compensator $\phi$,
\begin{align}
J_{\alpha \dalpha} = \frac{\delta S}{\delta H^{\dalpha \alpha}}~,\qquad
X = -\frac{1}{3} \frac{\delta S}{\delta \phi}~.
\end{align}

The supercurrent associated with non-minimal supergravity is\footnote{Actually,
there is a broader class of non-minimal supercurrents associated with the
ability to redefine $\Gamma$ in the linearized action \eqref{eq_transAdS}
as follows $\G \to \G+\l {\bar \cD}_\ad \cD_\a H^{\a\ad}$, for a complex parameter $\l$.
The choice implicitly made here corresponds to the simplest form of the linearized
action.}
\begin{align}\label{eq_nonminCurrent}
\BCD^\dalpha J_{\alpha \dalpha} = -\frac{1}{4} \BCD^2 \zeta_\alpha~,\qquad
\CD_{(\beta} \zeta_{\alpha)} = 0~.
\end{align}
where $\zeta_\alpha$ is calculated by
\begin{align}
\zeta_\alpha = \frac{\delta S}{\delta \psi^\alpha}~.
\end{align}
The superfield $\psi_\alpha$ denotes a prepotential for $\bar\Gamma$,
\begin{align}
\bar \Gamma = \CD^\alpha \psi_\alpha + \cdots
\end{align}
where the ellipsis denotes matter contributions.\footnote{In other words,
the first term $\CD^\alpha \psi_\alpha$ corresponds to a complex
linear superfield $\bar\Sigma$ obeying the homogeneous linearity
condition $(\CD^2 - 4 \bar R) \bar\Sigma = 0$. It is this piece
which we can vary independently of the matter fields.}
The constraint $\cD_{(\beta} \zeta_{\alpha)} = 0$ arises because
$\psi_\alpha$ possesses the gauge invariance $\delta \psi_\alpha = \cD^\beta \Omega_{\beta \alpha}$
for $\Omega_{\beta \alpha} = \Omega_{\alpha \beta}$.
Equivalently, if the action is written only in terms of $\Gamma$
and $\bar\Gamma$, then
\begin{align}
\zeta_\alpha = \frac{\delta S}{\delta \psi^\alpha}
	= -\cD_\alpha \frac{\delta S}{\delta \bar\Gamma}
	= \cD_\alpha (V + \ri \,U)
\end{align}
for real well-defined operators $V$ and $U$.\footnote{It is important to note that
in AdS the equation $\CD_{(\beta} \zeta_{\alpha)} = 0$ can \emph{always}
be solved by $\zeta_\alpha = \CD_\alpha \zeta$ for a complex \emph{globally-defined}
superfield which can be chosen as $\zeta = \CD^\gamma \zeta_\gamma / 4 \bar\mu$. This is a major
difference from Minkowski space where one cannot guarantee a globally-defined
$\zeta$ always exists.}

It is easy to check that the AdS supercurrents are related to
each other by well-defined improvement transformations. The construction is a simple
AdS generalization of that given in \cite{Kuzenko:2010ni} for the Minkowski case.
Beginning with the non-minimal supercurrent \eqref{eq_nonminCurrent},
one can construct
\begin{align}
\hat J_{\alpha \dalpha} = J_{\alpha \dalpha} + \frac{1}{6} [\cD_\alpha, \bar\cD_\dalpha] V
	- \cD_{\alpha \dalpha} U~,\qquad
\hat X := \frac{1}{12} (\BCD^2 - 4 \mu) (V - 3 \ri U)~.
\end{align}
It is a simple calculation to show that
\begin{align}
\bar \cD^\dalpha \hat J_{\alpha \dalpha}
	= \cD_\alpha \hat X~.
\end{align}
(The reverse relation may also be easily constructed in a similar way.)
It is easy to see that this improvement transformation matches the Minkowski
case which arises by setting $\mu = 0$. However, there is another purely AdS
feature which we can exploit:
\begin{align}
\frac{1}{4\mu} \BCD^2 \CD_\alpha X = \CD_\alpha X
\end{align}
when $X$ is chiral, $\BCD_\dalpha X = 0$. This allows us to absorb
\emph{any} contribution $\CD_\alpha X$ into the term $\zeta_\alpha$
via the redefinition
\begin{align}
\zeta_\alpha' = \zeta_\alpha - \frac{1}{\mu} \CD_\alpha X~.
\end{align}
What is remarkable about this transformation is that it requires
no assumptions about the global definition of the superfield $X$
itself; this transformation is globally defined so long as the original
trace multiplet is globally defined.

In Minkowski space, it has recently been argued that the
$\cS$-multiplet \cite{KS}
\begin{align}
\bar D^\dalpha \cS_{\alpha \dalpha} = \chi_\alpha + D_\alpha X~,\qquad
\bar D_\dalpha \chi_\alpha = \bar D_\dalpha X = D^\alpha \chi_\alpha - \bar D_\dalpha \bar\chi^\dalpha = 0~.
\end{align}
and its natural generalization given in \cite{Dumitrescu:2011iu} is the most
general (although it does not correspond to any known \emph{irreducible} $\N=1$ supergravity
formulation). Our interpretation of this statement is that 
that any other supercurrent which appears in a particular supersymmetric field
theory in Minkowski space may be related to the $\cS$-multiplet by a well-defined
improvement transformation. It is worth pointing out that the $\cS$-multiplet is extremely
natural in $\N=2$ supersymmetric field theories \cite{BK2011} as it is contained
within the $\N=2$ supercurrent which we constructed in \cite{Butter:2010sc}.
However, the extension of the $\cS$-multiplet to $\N=1$ AdS is unclear \cite{BK2011}.
As we have discussed there are only two irreducible supercurrent multiplets in AdS.

As a simple illustration of AdS supercurrents, we consider the case of the most general
non-linear sigma model in AdS,
\begin{align}
S = \fint K + \cint W_{\rm AdS} + \acint \bar W_{\rm AdS}
\end{align}
which can be rewritten using the AdS version of \eqref{eq_DtoFterm} as
\begin{align}
S = \fint \cK~,\qquad
\cK := K + \frac{W_{\rm AdS}}{\mu} + \frac{\bar W_{\rm AdS}}{\bar\mu}~.
\end{align}
The model is invariant under K\"ahler transformations of the form
\begin{align}
K \rightarrow K + F + \bar F~, \qquad
W_{\rm AdS} \rightarrow W_{\rm AdS} - \mu F
\end{align}
for which the combination $\cK$ is inert.
This action may be understood as the $\kappa\rightarrow 0$ limit of
the supergravity model
\begin{align}
S &= -\frac{3}{\kappa^2} \fint \phi \bar\phi \,{\rm e}^{-\kappa^2 K/3}
	+ \cint \phi^3 W
	+ \acint \bar\phi^3 \bar W
\end{align}
if we encode the AdS superpotential into the supergravity superpotential $W$
via
\begin{align}
W &= \frac{\mu}{\kappa^2} \exp(\kappa^2 W_{\rm AdS} / \mu)~.
\end{align}
This identification is useful since then the K\"ahler transformations have
exactly the same form in both the supergravity formulation and in the
AdS limit.

In the full supergravity background in the gauge where $\phi = 1$,
the supercurrent equation can be written
\begin{align}
\BCD^\dalpha J_{\alpha \dalpha} = \CD_\alpha X
\end{align}
where the derivatives $\CD_A$ are the usual Wess-Zumino
superspace derivatives. The real superfield $J_{\alpha \dalpha}$
and the chiral superfield $X$ are given respectively by
\begin{subequations}
\begin{align}
J_{\alpha \dalpha} &:=
	\frac{1}{\kappa^2} \re^{-\kappa^2 K/3} \left(
	G_{\alpha \dalpha} +
	\re^{-\kappa^2 K/6} [\CD_\alpha, \BCD_\dalpha] \re^{\kappa^2 K/6} 
	- \frac{\kappa^2}{2} K_{i \bar j} \CD_\alpha \phi^i \BCD_\dalpha \bar\phi^{\bar j} 
	\right) \eol
	&= 
	\re^{-\kappa^2 K/3} \left(\frac{1}{\kappa^2} G_{\alpha \dalpha} +
	\frac{1}{6} [\CD_\alpha, \BCD_\dalpha] K
	+ \frac{\kappa^2}{18} \CD_\alpha K \BCD_\dalpha K
	- \frac{1}{2} K_{i \bar j} \CD_\alpha \phi^i \BCD_\dalpha \bar\phi^{\bar j} \right)\\
X &:= -\frac{1}{4\kappa^2} (\BCD^2 - 4 R) \re^{-\kappa^2 K/3} - W \eol
	&=  \re^{-\kappa^2 K/3} \left(\frac{1}{\kappa^2} R
	+ \frac{1}{12} \BCD^2 K
	- \frac{\kappa^2}{36} \BCD_\dalpha K \BCD^\dalpha K \right)
	- W~.
\end{align}
\end{subequations}
This form involves operators which are not K\"ahler
invariant, but this is no surprise; in the frame $\phi=1$,
every K\"ahler transformation must be accompanied by a super-Weyl
transformation to restore invariance.
One is tempted to identify the first term of $J_{\alpha \dalpha}$
and of $X$ as the
``supergravity term'' and the rest as ``matter contributions;''
however, this is completely dependent on the way one defines
the supergeometry. For example, if we exchange Wess-Zumino
superspace geometry for K\"ahler superspace geometry, the
supercurrent equation takes the form
\begin{align}
\FBCD^\dalpha J_{\alpha \dalpha} = \FCD_\alpha X
\end{align}
where
\begin{align}
J_{\alpha \dalpha} := \frac{1}{\kappa^2} \FG_{\alpha \dalpha}
	- \frac{1}{2} K_{i \bar j} \FCD_\alpha \phi^i \FBCD_\dalpha \bar\phi^{\bar j}~,\qquad
X := \frac{1}{\kappa^2} \FR - \cW~.
\end{align}
Here $\FCD_A$ is the covariant derivative and $\FG_{\alpha \dalpha}$ and $\FR$ the torsion
tensors in the K\"ahler superspace formulation.
This naively has the Ferrara-Zumino form and
the first terms of both $J_{\alpha\dalpha}$ and $X$ involve what appear to
be purely supergravity terms. However, this is misleading as $\FR$ and $\FG$
obey the K\"ahler superspace constraint
\begin{align}
\FCD_\alpha \FR - \FBCD^\dalpha \FG_{\alpha \dalpha} = -\frac{\kappa^2}{4} (\FBCD^2 - 4 \FR) \FCD_\alpha K
\end{align}
which is decidedly a matter coupling. Making use of this identity, the
supercurrent equation can be recast in the form
\begin{align}\label{eq_Ksupercurrent}
-\frac{1}{2} \FBCD^\dalpha \FK_{\alpha \dalpha} = \FK_\alpha - \FCD_\alpha \cW
\end{align}
where
\begin{align}
\FK_{\alpha \dalpha} := K_{i \bar j} \FCD_\alpha \phi^i \FBCD_\dalpha \bar\phi^{\bar j}~,\quad
\FK_\alpha := -\frac{1}{4} (\FBCD^2 - 4 \FR) \FCD_\alpha K~.
\end{align}
This equation holds on-shell due to the matter equation of motion
\begin{align}
\frac{1}{4} \FBCD^2 K_i = \nabla_i \cW
\end{align}
where in the Hermitian basis
\begin{align}
\nabla_i \cW := \re^{\kappa^2 (K/2 + W_{\rm AdS} / \mu)} \partial_i \left(W_{\rm AdS} + \mu K\right)~\qquad
\textrm{(Hermitian basis)}~.
\end{align}
The supercurrent \eqref{eq_Ksupercurrent} has the form of the $\cS$-multiplet but again
this is somewhat misleading since $\cW$ is covariant with respect to K\"ahler
transformations and the derivatives carry a K\"ahler connection.
When we take the $\kappa \rightarrow 0$ limit, we recover (being careful to include
the contribution from the $U(1)_K$ connection)
\begin{align}
- \frac{1}{2} \BCD^\dalpha K_{\alpha \dalpha} &= K_\alpha - \CD_\alpha (W_{\rm AdS} + \mu K)
\end{align}
where
\begin{align}
K_{\alpha \dalpha} := K_{i \bar j} \CD_\alpha \phi^i \BCD_\dalpha \bar\phi^{\bar j}~,\qquad
K_\alpha := -\frac{1}{4} (\BCD^2 - 4 \mu) \CD_\alpha K
\end{align}
with the derivatives $\CD_A$ obeying the AdS superspace algebra \eqref{N=1-AdS-algebra}.
This is still not quite the $\cS$-multiplet due to the non-chiral term $\mu K$
appearing in the supercurrent equation.\footnote{This expression may still be
interpreted as the $\cS$-multiplet if one thinks of the $\mu K$ term as arising from
the connection for the nonlinear K\"ahler transformation of $W_{\rm AdS}$, but this
seems to violate the spirit of the $\cS$-multiplet.}
On the other hand, we observe that the equation may be rewritten in the form
\begin{align}
- \frac{1}{2} \BCD^\dalpha K_{\alpha \dalpha} &=
	- \frac{1}{4} \BCD^2 \CD_\alpha K
	- \CD_\alpha W_{\rm AdS} \eol
	&= 
	-\frac{1}{4} \BCD^2 \CD_\alpha
	\left(K + \frac{W_{\rm AdS}}{\mu} + \frac{\bar W_{\rm AdS}}{\bar \mu} \right)
	= -\frac{1}{4} \BCD^2 \CD_\alpha \cK
\end{align}
where $\cK$ is the globally defined Lagrangian for the general
non-linear sigma model in AdS.
This has precisely the form of the non-minimal $(n=-1)$ supercurrent
conservation equation and one may check by an explicit calculation that it
holds due to the matter equation of motion in AdS,
\begin{align}
\frac{1}{4} \BCD^2 K_i = \partial_i (W_{\rm AdS} + \mu K) = \mu \partial_i \cK \implies
\frac{1}{4} (\BCD^2 - 4 \mu) \cK_i = 0~.
\end{align}

In light of these considerations we propose the following supercurrent
formula as the general supercurrent in AdS
\begin{align}\label{eq_AdScurrent}
\BCD^\dalpha J_{\alpha \dalpha} = \CD_\alpha X -\frac{1}{4} \BCD^2 \zeta_\alpha~, \qquad
\BCD_\dalpha X = \CD_{(\beta} \zeta_{\alpha)} = 0~.
\end{align}
Each of the trace multiplets, $X$ and $\zeta_\alpha$,
may be identified with a specific linearized action in AdS,
which due to the results of this paper, can \emph{both} be
extended to a full nonlinear action. While the inclusion of both
trace multiplet contributions seems to imply this is a $24+24$
component supercurrent, we must keep in mind the symmetry
\begin{align}
X \rightarrow X + \mu \Lambda~,\qquad
\zeta_\alpha \rightarrow \zeta_\alpha + \CD_\alpha \Lambda~,\qquad
\BCD_\dalpha \Lambda = 0
\end{align}
which we exploited above. This reduces the supercurrent to $20+20$
components.

It is interesting that the AdS supercurrent \eqref{eq_AdScurrent}
reduces in the naive Minkowski limit to a $20+20$
supercurrent of the form\footnote{This supercurrent was considered in a slightly
different form in \cite{Magro:2001aj}.},
\begin{subequations}
\begin{gather}
\bar D^\dalpha J_{\alpha \dalpha} = D_\alpha X + \chi_\alpha + \ri \,\eta_\alpha \\
\bar D_\dalpha X = \bar D_\alpha \chi_\alpha = \bar D_\dalpha \eta_\alpha = 
D^\alpha \chi_\alpha - \bar D_\dalpha \bar\chi^\dalpha =
D^\alpha \eta_\alpha - \bar D_\dalpha \bar\eta^\dalpha = 0
\end{gather}
\end{subequations}
which arises from the consideration of all the possible models of
linearized supergravity categorized in recent years
(see \cite{Kuzenko:2010am, Kuzenko:2010ni} and the references therein).
It was argued in \cite{Dumitrescu:2011iu} that this more general
supercurrent can always be improved to the $\cS$-multiplet. The closest
analogy for the proposed AdS supercurrent \eqref{eq_AdScurrent} would
be the statement that $\zeta_\alpha$ may always be improved to
$\zeta_\alpha = \CD_\alpha U$ for \emph{real} superfield $U$.
If so, then the supercurrent \eqref{eq_AdScurrent} would be
reduced to $16+16$ components in AdS and naturally approach the
$\cS$-multiplet form in the naive Minkowski limit.

\noindent
{\bf Acknowledgements:}\\
This work is supported in part by the Australian Research Council 
and by a UWA Research Development Award.

\appendix

\section{The massless superspin-3/2 multiplet in AdS}
Off-shell massless higher-superspin multiplets were constructed in AdS
superspace  in \cite{KS94}. For the case of
half-integer superspins, they are characterized by a family of
dually equivalent models beginning at superspin-3/2. 
The so-called longitudinal formulation of the free
massless superspin-3/2 multiplet in AdS can be written
\begin{align}\label{eq_longAds}
S_{(3/2)}^{||} &= - \fint \Big\{
      \frac{1}{16} H^{\dalpha \alpha} \cD^\beta (\bar\cD^2 - 4R) \cD_\beta H_{\alpha \dalpha}
	- \frac{1}{48} ([\cD_\alpha, \bar\cD_\dalpha] H^{\dalpha \alpha})^2 
     \eol & \quad
     + \frac{1}{4} (\cD_{\alpha \dalpha} H^{\dalpha \alpha})^2
     + \frac{1}{4} R \bar R H^{\dalpha \alpha} H_{\alpha \dalpha}
     + \ri H^{\dalpha \alpha} \cD_{\alpha \dalpha} (\f - \bar \f)
     + 3 ( \f \bar\f  -  \f^2 -  \bar\f^2) \Big\}~.
\end{align}
The superfield $H_{\alpha \dalpha}$ is real while $\f$ is covariantly chiral,
\begin{align}
\bar\cD_\dalpha \f = 0~.
\end{align}
The action possesses the gauge invariance
\begin{align}
\delta H_{\alpha \dalpha} = \cD_\alpha \bar L_\dalpha - \bar\cD_\dalpha L_\alpha~,\qquad
\delta \f = -\frac{1}{12} (\bar\cD^2 - 4 R) \cD^\alpha L_\alpha~.
\end{align}
This is exactly the linearized action of old minimal supergravity with
a cosmological term.

This model may be dualized to one involving a complex linear multiplet
$\Gamma$, obeying
\begin{align}
(\bar\cD^2 - 4 R) \Gamma = 0~.
\end{align}
The dual action, which is called the transverse formulation in the context
of \cite{KS94}, is
\begin{align}\label{eq_transAdS}
S_{(3/2)}^\perp &=- \fint \Big\{
	\frac{1}{16} H^{\dalpha \alpha} \cD^\beta (\bar\cD^2 - 4R) \cD_\beta H_{\alpha \dalpha} 
	+ \frac{1}{4} R \bar R H^{\dalpha \alpha} H_{\alpha \dalpha}
	\eol & \quad
	+ \frac{1}{2} H^{\a\ad} (\cD_\a {\bar \cD}_\ad \G - {\bar \cD}_\ad \cD_\a {\bar \G} ) 
	+ \bar \G \G + \G^2 +{\bar \G}^2 \Big\} ~,
\end{align}
with the gauge invariance
\begin{align}
\delta H_{\alpha \dalpha} = \cD_\alpha \bar L_\dalpha - \bar\cD_\dalpha L_\alpha~,\qquad
\delta \Gamma = -\frac{1}{4} \bar\cD_\dalpha \cD^2 \bar L^\dalpha~.
\end{align}

\footnotesize{

}

\end{document}